\documentstyle [12pt,amsfonts] {article}
\input epsf

\newcommand{\beq}{\begin{equation}}
\newcommand{\eeq}{\end{equation}}
\newcommand{\beqs}{\begin{eqnarray}}
\newcommand{\eeqs}{\end{eqnarray}}

\catcode`@=11
\@addtoreset{equation}{section}
\@addtoreset{equation}{subsection}
\def\theequation{\ifnum\value{section}=0 \arabic{equation}\ignorespaces
\else \ifnum\value{section}=-1 A.\arabic{equation}\ignorespaces
\else \ifnum\value{subsection}=0 \thesection.\arabic{equation}\ignorespaces
\else \thesection.\arabic{subsection}.\arabic{equation}\ignorespaces
                           \fi
                      \fi
                 \fi}
\catcode`@=12

\begin{document}

\def\thefootnote{\fnsymbol{footnote}}

\baselineskip 5.0mm

\vspace{4mm}

\begin{center}

{\Large \bf Structural Properties of Potts Model Partition Functions and
Chromatic Polynomials for Lattice Strips} 

\vspace{8mm}

\setcounter{footnote}{0}
Shu-Chiuan Chang$^{(a)}$\footnote{email: shu-chiuan.chang@sunysb.edu} and
\setcounter{footnote}{6}
Robert Shrock$^{(a,b)}$\footnote{(a): permanent address;
email: robert.shrock@sunysb.edu}

\vspace{6mm}

(a) \ C. N. Yang Institute for Theoretical Physics  \\
State University of New York       \\
Stony Brook, N. Y. 11794-3840  \\

(b) \ Physics Department \\
Brookhaven National Laboratory \\
Upton, NY  11973-5000

\vspace{10mm}

{\bf Abstract}
\end{center}

The $q$-state Potts model partition function (equivalent to the Tutte 
polynomial) for a lattice strip of fixed width $L_y$ and arbitrary length $L_x$
has the form
$Z(G,q,v)=\sum_{j=1}^{N_{Z,G,\lambda}}c_{Z,G,j}(\lambda_{Z,G,j})^{L_x}$, where
$v$ is a temperature-dependent variable.  The special case of the
zero-temperature antiferromagnet ($v=-1$) is the chromatic polynomial
$P(G,q)$.  Using coloring and transfer matrix methods, we give general 
formulas for $C_{X,G}=\sum_{j=1}^{N_{X,G,\lambda}}c_{X,G,j}$ for
$X=Z,P$ on cyclic and M\"obius strip graphs of the square and triangular
lattice.  Combining these with a general expression for the (unique)
coefficient $c_{Z,G,j}$ of degree $d$ in $q$:
$c^{(d)}=U_{2d}(\frac{\sqrt{q}}{2})$, where $U_n(x)$ is the Chebyshev
polynomial of the second kind, we determine 
the number of $\lambda_{Z,G,j}$'s with coefficient $c^{(d)}$ in 
$Z(G,q,v)$ for these cyclic strips of width 
$L_y$ to be $n_Z(L_y,d)=(2d+1)(L_y+d+1)^{-1} 
{2L_y \choose L_y-d}$ for $0 \le d \le L_y$ and zero otherwise.  For both
cyclic and M\"obius strips of these lattices, the total number of distinct 
eigenvalues $\lambda_{Z,G,j}$ is calculated to be 
$N_{Z,L_y,\lambda}={2L_y \choose L_y}$.  
Results are also presented for the analogous numbers
$n_P(L_y,d)$ and $N_{P,L_y,\lambda}$ for $P(G,q)$.  We find that 
$n_P(L_y,0)=n_P(L_y-1,1)=M_{L_y-1}$ (Motzkin number), 
$n_Z(L_y,0)=C_{L_y}$ (the Catalan number), and give an exact expression for 
$N_{P,L_y,\lambda}$.  Our results for $N_{Z,L_y,\lambda}$ and
$N_{P,L_y,\lambda}$ apply for both the cyclic and M\"obius strips of both the
square and triangular lattices; we also point out the interesting relations 
$N_{Z,L_y,\lambda}=2N_{DA,tri,L_y}$ and 
$N_{P,L_y,\lambda}=2N_{DA,sq,L_y}$,
where $N_{DA,\Lambda,n}$ denotes the number of directed lattice animals on the
lattice $\Lambda$.  We find the asymptotic growths $N_{Z,L_y,\lambda}
\sim L_y^{-1/2} \ 4^{L_y}$ and $N_{P,L_y,\lambda} \sim L_y^{-1/2} \ 3^{L_y}$ as
$L_y \to \infty$.  Some general geometric identities for Potts model partition
functions are also presented.

\vspace{16mm}

\pagestyle{empty}
\newpage

\pagestyle{plain}
\pagenumbering{arabic}
\renewcommand{\thefootnote}{\arabic{footnote}}
\setcounter{footnote}{0}

\section{Introduction}

The $q$-state Potts model has served as a valuable model for the study of phase
transitions and critical phenomena \cite{potts,wurev}.  On a lattice, or, more
generally, on a (connected) graph $G$, at temperature $T$, this model is
defined by the partition function
\beq
Z(G,q,v) = \sum_{ \{ \sigma_n \} } e^{-\beta {\cal H}}
\label{zfun}
\eeq
with the (zero-field) Hamiltonian
\beq
{\cal H} = -J \sum_{\langle i j \rangle} \delta_{\sigma_i \sigma_j}
\label{ham}
\eeq
where $\sigma_i=1,...,q$ are the spin variables on each vertex $i \in G$;
$\beta = (k_BT)^{-1}$; and $\langle i j \rangle$ denotes pairs of adjacent
vertices.  The graph $G=G(V,E)$ is defined by its vertex set $V$ and its edge
set $E$; we denote the number of vertices of $G$ as $n=n(G)=|V|$ and the
number of edges of $G$ as $e(G)=|E|$.  We use the notation
\beq
K = \beta J \ , \quad a = u^{-1} = e^K \ , \quad v = a-1
\label{kdef}
\eeq
so that the physical ranges are (i) $a \ge 1$, i.e., $v \ge 0$ corresponding to
$\infty \ge T \ge 0$ for the Potts ferromagnet, and (ii) $0 \le a \le 1$,
i.e., $-1 \le v \le 0$, corresponding to $0 \le T \le \infty$ for the Potts
antiferromagnet.  One defines the (reduced) free energy per site $f=-\beta F$,
where $F$ is the actual free energy, via
\beq
f(\{G\},q,v) = \lim_{n \to \infty} \ln [ Z(G,q,v)^{1/n}]  \ .
\label{ef}
\eeq
where we use the symbol $\{G\}$ to denote $\lim_{n \to \infty}G$ for a given
family of graphs.

Let $G^\prime=(V,E^\prime)$ be a spanning subgraph of $G$, i.e. a subgraph
having the same vertex set $V$ and an edge set $E^\prime \subseteq E$. Then
$Z(G,q,v)$ can be written as the sum \cite{birk}-\cite{fk}
\beqs
Z(G,q,v) & = & \sum_{G^\prime \subseteq G} q^{k(G^\prime)}v^{e(G^\prime)}
\label{cluster} \cr\cr\cr
& = & \sum_{r=k(G)}^{n(G)}\sum_{s=0}^{e(G)}z_{rs} q^r v^s
\label{zpol}
\eeqs 
where $k(G^\prime)$ denotes the number of connected components of
$G^\prime$ and $z_{rs} \ge 0$.  Since we only consider connected graphs $G$, we
have $k(G)=1$. The formula (\ref{cluster}) shows that $Z(G,q,v)$ is a
polynomial in $q$ and $v$ and enables one to generalize $q$ from ${\mathbb
Z}_+$ to ${\mathbb R}_+$ and, indeed, to ${\mathbb C}$.  The Potts model
partition function $Z(G,q,v)$ on a graph $G$ is essentially equivalent to the
Tutte polynomial \cite{tutte1}-\cite{tutte3} $T(G,x,y)$ and Whitney rank
polynomial $R(G,\xi,\eta)$ \cite{wurev},\cite{whit},\cite{bbook}
for this graph, as discussed in the appendix.

One special case of the Potts model partition function that is of particular 
interest is the zero-temperature limit of the Potts antiferromagnet (AF). For 
sufficiently large $q$, on a given lattice or graph $G$, this
exhibits nonzero ground state entropy (without frustration).
This is equivalent to a ground
state degeneracy per site (vertex), $W > 1$, since $S_0 = k_B \ln W$.  The
$T=0$ (i.e., $v=-1$) partition function of the above-mentioned $q$-state Potts
antiferromagnet (PAF) on $G$ satisfies
\beq 
Z(G,q,-1)=P(G,q)
\label{zp}
\eeq
where $P(G,q)$ is the chromatic polynomial (in $q$) expressing the number
of ways of coloring the vertices of the graph $G$ with $q$ colors such that no
two adjacent vertices have the same color \cite{birk,bbook,rrev,rtrev}. This
is termed a proper vertex coloring of $G$. The
minimum number of colors necessary for this coloring is the chromatic number of
$G$, denoted $\chi(G)$.  Thus\footnote{\footnotesize{At certain special
points $q_s$ (typically $q_s=0,1,.., \chi(G)$), one has the noncommutativity of
limits $\lim_{q \to q_s} \lim_{n \to \infty} P(G,q)^{1/n} \ne \lim_{n \to
\infty} \lim_{q \to q_s}P(G,q)^{1/n}$, and hence it is necessary to specify the
order of the limits in the definition of $W(\{G\},q_s)$ \cite{w}.}}
\beq
W(\{G\},q) = \lim_{n \to \infty} P(G,q)^{1/n} \ . 
\label{w}
\eeq
A second special case of the Potts model partition function is for infinite
temperature, i.e., $K=0$ or equivalently $v=0$.  In this case, as is clear from
either (\ref{zfun}) with (\ref{ham}) or from (\ref{cluster}), $Z$ reduces to
the single term 
\beq
Z(G,q,0)=q^{n(G)} \ .
\label{zvzero}
\eeq
A third special case is the Potts ferromagnet in the limit of zero temperature,
$T \to 0$, i.e., $v \to \infty$.

Here we shall consider families of graphs that are cyclic strips of the square
and triangular lattices, taken to be oriented horizontally, with free
transverse boundary conditions, denoted $FBC_y$, and periodic longitudinal
boundary conditions, denoted $PBC_x$.  A given strip of the triangular 
lattice may be visualized by starting with the corresponding strip of the 
square lattice and adding diagonal bonds joining, say, the upper left to 
lower right vertices of each square.  The length and width are taken to be
$L_x$ and $L_y$ vertices, and the total number of vertices is $n=L_xL_y$.
We shall also make some comments about the analogous
strips with twisted periodic longitudinal, i.e., M\"obius, boundary conditions,
denoted $TPBC_x$ or $Mb$.  These various families of lattice strip graphs are
examples of recursive families of graphs, in the sense that a strip of length
$L_x+1$ is constructed by the addition of a given subgraph (here, a transverse
layer of the strip) to the strip of length $L_x$.  As derived in \cite{bcc,a},
using a type of transfer matrix argument, a general form for the Potts model
partition function for the strip graphs considered here, or more generally, for
a recursively defined graph comprised of $L_x=m$ repeated subunits (transverse
layers of the strip here), is 
\beq 
Z(G,q,v) =
\sum_{j=1}^{N_{Z,G,\lambda}} c_{Z,G,j}(\lambda_{Z,G,j})^m
\label{zgsum}
\eeq 
where $\lambda_{Z,G,j}$ is a function of $q$ and $v$, and both
$\lambda_{Z,G,j}$ and the coefficients $c_{Z,G,j}$ are independent of
$L_x$.  The coefficients $c_{Z,G,j}$ are functions only of $q$ and, indeed, are
polynomials, for the cyclic strip graphs of interest here. (We shall comment on
other types of strip graphs below.)

 From (\ref{zp}), it also follows that the chromatic polynomial has the
same structure, 
\beq
P(G,q) = \sum_{j=1}^{N_{P,G,\lambda}} c_{P,G,j}(\lambda_{P,G,j})^m
\label{pgsum}
\eeq 
where $N_{P,G,\lambda}$ depends on $G$; $c_{P,G,j}$ and 
$\lambda_{P,G,j}$ are independent of $L_x=m$; and 
\beq
N_{P,G,\lambda} \le N_{Z,G,\lambda} \ . 
\label{nnineq}
\eeq
As will be shown below, for the strips under consideration here, this 
inequality is realized as an equality for the circuit graph $C_n$ and as a 
strict inequality for widths $L_y \ge 2$ (see Tables \ref{npctablecyc} and 
\ref{ntctablecyc}).  For the same reason, it follows that when one sets $v=-1$
in $Z(G,q,v)$, i.e. specializes to the $T=0$ Potts antiferromagnet, a certain
number $N_{Z,G,\lambda}-N_{P,G,\lambda}$ of the $\lambda_{Z,G,j}$'s 
vanish and the remaining, nonvanishing $\lambda_{Z,G,j}(q,v)$'s are precisely 
those occurring in $P(G,q)$:
\beq
{\rm If} \ \ \lambda_{Z,G,j}(q,v=-1) \ne 0 \ \ 
{\rm then} \ \ \lambda_{Z,G,j}(q,-1) = \lambda_{P,G,j}(q) \ . 
\label{lamlameq}
\eeq 
Previous calculations of chromatic polynomials for recursive families of
graphs of arbitrary length include \cite{bds}-\cite{ss}; we shall also use the
results of calculations of Potts model partition functions (Tutte polynomials)
for cyclic and M\"obius strip graphs of arbitrary length in
\cite{bcc,a,ta}.  Some relevant results 
on transfer matrices and the Temperley-Lieb algebra are in \cite{baxter}, 
\cite{templieb}--\cite{mbook}. 

Several basic questions about the structure of $Z(G,q,v)$ and $P(G,q)$ for
these cyclic strip graphs are the following: 

\begin{enumerate}

\item 

Can one obtain a general formula for the coefficients $c_{Z,G,j}$ and
$c_{P,G,j}$ in (\ref{zgsum}) and (\ref{pgsum})? 

\item

What are the respective sums of the coefficients 
$c_{Z,G,j}$ and $c_{P,G,j}$ in (\ref{zgsum}) and (\ref{pgsum})? 

\item

For a given cyclic strip $G$, how many $\lambda_{Z,G,j}$'s in (\ref{zgsum}) 
have a particular coefficient, and how many $\lambda_{P,G,j}$'s in
(\ref{pgsum}) have this coefficient? 

\item

What are the total numbers of different terms $N_{Z,G,\lambda}$ and
$N_{P,G,\lambda}$ in (\ref{zgsum}) and (\ref{pgsum})? 

\end{enumerate}

In this paper, we shall obtain a general formula for the coefficients
$c_{Z,G,j}$ and $c_{P,G,j}$, an answer to the second pair of
questions, and, based on these results, answers to the remaining
two pairs of questions.  We shall also comment on other types of lattice
strips.

\vspace{8mm}

The chromatic polynomial, can be calculated by various methods, including
iterative application of the deletion-contraction theorem (e.g., \cite{bbook}),
a certain matrix technique \cite{bm}, a generating function method
\cite{strip,hs}, or a coloring (compatibility) matrix method \cite{b,matmeth}.
Similarly, the Potts model partition function or equivalent Tutte polynomial 
can be calculated by iterative application of the generalized 
deletion-contraction theorem \cite{bbook} or by transfer matrix methods
including a relation with the Temperley-Lieb algebra \cite{baxter,baxbook,
mbook}.  
Of course these different methods can be used to provide cross-checks; for
example, iterated deletion-contraction and transfer matrix methods were used to
check each other in \cite{a,ta} and deletion-contraction and coloring matrix
methods were used together in works such as \cite{tk}. 
The last, the coloring matrix method, is also useful for deriving rigorous
upper and lower bounds on $W(\{G\},q)$ \cite{b},\cite{ww},\cite{wn}.  In this
method, one first selects a transverse slice of the strip, denoted ${\cal
L}_{L_y}$.  For our cyclic graphs, this is simply a line graph with $L_y$
vertices (oriented vertically, given that we orient the long direction of the
strip horizontally).  Denote an allowed $q$-coloring of this path as $c({\cal
L}_{L_y})$.  The number of allowed colorings of the path ${\cal
L}_{L_y}$ is ${\cal N} = P({\cal L}_{L_y},q)$.  For our chromatic polynomials
of cyclic strips of the square and triangular lattices, $P({\cal
L}_n,q)=P(T_{L_y},q)$, where $T_n$ is the tree graph with $n$ vertices and 
\beq
P(T_n,q)=q(q-1)^{n-1} \ . 
\label{pt}
\eeq 
Now focus on two adjacent paths ${\cal L}_{L_y}$ and ${\cal L}_{L_y}^\prime$.
Define compatible $q$-colorings of these paths as colorings
such that no two adjacent vertices $v \in {\cal L}_{L_y}$ and $v' \in {\cal
L}_{L_y}^\prime$ (i.e. vertices connected by an edge = bond of the lattice
strip graph) have the same color.  
One can then associate with this pair of paths an ${\cal N} \times
{\cal N}$ dimensional symmetric matrix ${\cal T}$ with entries 
${\cal T}_{c({\cal L}_{L_y}),c({\cal L}_{L_y}^\prime)}=1$ or 0 if the 
$q$-colorings of ${\cal L}_{L_y}$ and ${\cal L}_{L_y}^\prime$ 
are or are not compatible, respectively. Then the chromatic polynomial
of the cyclic strip of the lattice $\Lambda$, taken here to be square ($sq$) or
triangular ($tri$), is (with $L_x=m$) given by 
\beq 
P(\Lambda, L_y \times L_x, FBC_y,PBC_x,q)=Tr({\cal T}^m) \ .
\label{pttrace}
\eeq
Since ${\cal T}$ is a
symmetric real matrix (indeed, composed only of 0's and 1's, although we do not
need this here), it can be diagonalized by an orthogonal transformation, so
that the above trace is $Tr({\cal T}^m)=\sum_j (\lambda_{P,G,j})^m$, where the
$\lambda_{P,G,j}$'s are the eigenvalues of ${\cal T}$.  In the context of
the
$q$-coloring problem, if one denotes the multiplicity of the $j$'th distinct
eigenvalue by $c_{P,G,j}$, one obtains the formula (\ref{pgsum}).  That is,
from the coloring matrix viewpoint, $c_{P,G,j}$ is the dimension of the
invariant subspace in the full ${\cal N}$-dimensional space of coloring 
configurations of the transverse slice of the strip 
corresponding to the eigenvalue $\lambda_{P,G,j}$. 
For the full temperature-dependent Potts model partition function, one can
define an analogous matrix, ${\cal T}_Z$ whose entries, rather than being 0 and
1, are appropriate Boltzmann weights for the given spin configurations on the
successive transverse slices ${\cal L}_{L_y}$ and ${\cal L}_{L_y}^\prime$.
Then this partition function is (with $L_x=m$)
\beq
Z(\Lambda, L_y \times L_x, FBC_y,PBC_x,q,v)=Tr[({\cal T}_Z)^m]
\label{zttrace}
\eeq 
which was used in \cite{bcc,a} to derive the formula (\ref{zgsum}).  Note that
in this case all colorings, including those that yield the same color on
adjacent vertices, are allowed.  One can obtain these formulas (\ref{pttrace})
and (\ref{zttrace}) 
for sufficiently large integral $q$ that the multiplicities
$c_{P,G,j}$ and $c_{Z,G,j}$ are positive integers; in both of these cases, the
fact that the coefficients are multiplicities shows that for sufficiently large
positive integer $q$ they are also positive integers.  In the case of the
chromatic polynomial, it is obvious that $c_{P,G,j}$ can depend only on $q$; in
the case of the full Potts model partition function, for the cyclic strip
graphs considered here, this also follows from (\ref{zttrace}) and
(\ref{zgsum}), since the multiplicities of the eigenvalues cannot depend on the
variable parameter $v \in (-1,\infty)$.  

Note that, while the coefficients $c_{P,G,j}$ and $c_{Z,G,j}$ can be obtained
as multiplicities of the distinct eigenvalues $\lambda_{P,G,j}$ and
$\lambda_{Z,G,j}$ for sufficiently large integer $q$, when one considers
positive $q < 4$, they may be zero or negative (see
eqs. (\ref{cdq0})-(\ref{cdq4}) below).  More generally, while they play the
role of eigenvalue multiplicities for sufficiently large integer $q$, the
domain of their definition may be generalized via (\ref{cluster}) to $q \in
{\mathbb R}_+$ or, indeed, to $q \in {\mathbb C}$.  Some properties of
determinants of coloring matrices for various strip graphs are given in the
appendix. 

The dimension of the space of coloring configurations, ${\cal N}$, is equal to
the sum of the multiplicities of each distinct eigenvalue, i.e., the sum of the
dimensions of the invariant subspaces corresponding to each of these distinct
eigenvalues.  For the chromatic polynomial, this is ${\cal N}$, which is equal
to the sum 
\beq
C_{P,G}=\sum_{j=1}^{N_{P,G,\lambda}} c_{P,G,j}
\label{cpsum}
\eeq
while for the full Potts model partition function, we shall denote it as 
\beq
C_{Z,G}=\sum_{j=1}^{N_{Z,G,\lambda}} c_{Z,G,j} \ . 
\label{czsum}
\eeq
General results for these sums will be given below for the strips of interest.

Before proceeding, we note that for the chromatic polynomial, since two
identical color configurations on ${\cal L}_{L_y}$ and ${\cal L^\prime}_{L_y}$
are incompatible, the diagonal elements of the coloring matrix are zero and 
hence its trace is zero: 
\beq 
Tr({\cal T}(G,q))=0 \quad {\rm for} \ \ P(G,q) \ . 
\label{ttracezero}
\eeq
Since the coefficients $c_{P,G,j}$ are just the multiplicities of the
given
eigenvalues $\lambda_{P,G,j}$, we can write this as the sum over distinct 
$\lambda_{P,G,j}$'s: 
\beq
\sum_{j=1}^{N_{P,G,\lambda}} c_{P,G,j}\lambda_{P,G,j}=0 \ . 
\label{clamzero}
\eeq
In the following, since we shall present results that are applicable to cyclic
strips of both the square and triangular lattices, we shall often leave the
lattice type $\Lambda$ and the boundary conditions $(FBC_y,PBC_x)$ implicit 
in the notation where they are obvious.  The fact that these results will be
applicable to cyclic strips of both the square and triangular lattices depends
on the property that (having expressed the strip of the triangular lattice as a
strip of the square lattice with diagonal bonds added to each square, as
described above) the transverse slices, i.e., line graphs of $L_y$ vertices, 
of both the cyclic square-lattice and triangular-lattice strips are
identical.\footnote{\footnotesize{Although the transverse slices are identical
for the cyclic strips of the square and triangular lattice, they are different
for other lattices, such as the honeycomb = brick lattice, and this 
difference was also significant for the earlier use of coloring matrix 
methods to obtain rigorous bounds on the Potts model ground state degeneracy 
per site, $W$ \cite{b}-\cite{wn}.}}  Similarly, for quantities that are
independent of some part of $G$, such as $c_{Z,G,j}$ and $\lambda_{Z,G,j}$,
which are independent of $L_x$, this will be incorporated in the notation.

\section{Coefficients in Potts Model Partition Functions for Cyclic Lattice
Strips}

In this section we address the first pair of questions posed in the
Introduction.  From our exact solutions of the chromatic polynomials and, more
generally, of the full temperature-dependent Potts model partition functions on
cyclic strips of the square and triangular lattice, we have found that the 
coefficients $c_{Z,G,j}$ and $c_{P,G,j}$ are polynomials in $q$.  Further, we 
have found that for a given cyclic strip of the square or triangular lattice
with width $L_y$, the coefficients $c_{P,G,j}$ and $c_{Z,G,j}$ are of a limited
set; only one type of polynomial of each degree in $q$ 
occurs\footnote{\footnotesize{From
the studies of M\"obius strips of the square lattice \cite{bds,wcy,pm,s4}, we 
find that the coefficients are again polynomials, of the form $\pm c^{(d)}$; 
however, for M\"obius strips of the triangular lattice, the coefficients are
not, in general, polynomials, but algebraic functions of $q$, as was shown by
the exact solutions for the chromatic polynomial in \cite{wcy} and for the 
full Potts model partition in \cite{ta} for the $L_y=2$ M\"obius strip.
For strips with torus boundary conditions, exact solutions \cite{tk,t} show
that the coefficients are polynomials of degree $d$ in $q$, but it is not, in 
general, the case that there is a unique coefficient of degree $d$ and they 
are not, in general, of the form $c^{(d)}$. We shall comment further on this 
below.}}, and, denoting
this as $c^{(d)}$, there are coefficients of the form $c^{(d)}$ with $0 \le d
\le L_y$.  We infer the following general formula for $c^{(d)}$, i.e. for the 
multiplicity of the corresponding eigenvalue of ${\cal T}_Z$: 
\beq 
c^{(d)}=U_{2d}\Bigl (\frac{\sqrt{q}}{2} \Bigr )
\label{cd}
\eeq
where $U_n(x)$ is the Chebyshev polynomial of the second kind, defined
by (e.g. \cite{gr,riordan})
\beqs
U_n(x) & = & \frac{1}{\sqrt{1-x^2}}Im \biggl [ (x+i\sqrt{1-x^2} \ )^{n+1} ]
\cr\cr & = & \sum_{j=0}^{[\frac{n}{2}]} (-1)^j {n-j \choose j} (2x)^{n-2j}
\label{undef}
\eeqs
where in eq. (\ref{undef}) and similar equations below, the notation 
$[\frac{n}{2}]$ in the upper limit on the summand means the integral part
of $\frac{n}{2}$.  In the first line of (\ref{undef}), $x$ must be real, but 
the polynomial in the second line of (\ref{undef}) constitutes a 
definition of $U_n(x)$ for complex $x$. 
The first few of these coefficients are, with $x=\frac{\sqrt{q}}{2}$, 
\beq
c^{(0)}=U_0(x)=1
\label{cd0}
\eeq

\beqs
c^{(1)} & = & U_2(x)=4x^2-1 \cr\cr
        & = & q-1
\label{cd1}
\eeqs

\beqs
c^{(2)} & = & U_4(x)=16x^4-12x^2+1 \cr\cr
        & = & q^2-3q+1
\label{cd2}
\eeqs

\beqs
c^{(3)} & = & U_6(x)=64x^6-80x^4+24x^2-1 \cr\cr
        & = & q^3-5q^2+6q-1 
\label{cd3}
\eeqs
\beqs
c^{(4)} & = & U_8(x)=256x^8-448x^6+240x^4-40x^2+1 \cr\cr
        & = & q^4-7q^3+15q^2-10q+1 \cr\cr
        & = & (q-1)(q^3-6q^2+9q-1)
\label{cd4}
\eeqs
and so forth for higher $d$.  We have found, as an equivalent formula 
\cite{t,s4} 
\beq
c^{(d)}= \prod_{k=1}^d (q - q_{d,k}) 
\label{cdprod}
\eeq
where 
\beq
q_{d,k} \equiv 2+2\cos \Bigl ( \frac{2\pi k}{2d+1} \Bigr )
 = 4\cos^2 \Bigl ( \frac{\pi k}{2d+1} \Bigr ) \ , \quad {\rm for} \quad 
k=1,2,...d \ . 
\label{cdzeros}
\eeq 
Note that the apparent square root in (\ref{cd}) is absent in the actual
formula (\ref{cd}) for the $c^{(d)}$'s, which applies for arbitrary complex 
$q$. One way to derive (\ref{cd}) for cyclic strips is to use the fact that the
partition function is given by the trace (\ref{zttrace}), which, with
(\ref{zgsum}) makes clear the role of the coefficients $c_{Z,G,j}$ as 
multiplicities of eigenvalues of the transfer matrix ${\cal T}_Z$; then, one
utilizes the connection with the Temperley-Lieb algebra to infer that these
multiplicities are given by $c^{(d)}$ (for sufficiently large integer $q$ where
the $c^{(d)}$ are positive integers), and finally, one continues this result to
general complex $q$.  Related discussions of dimensions of invariant subspaces
of operators in the Temperley-Lieb algebra are given in
\cite{martin},\cite{mbook}.  The form (\ref{cdprod}) is also related to the
property that each coefficient $c^{(d)}$ has a simple zero at the
Tutte-Beraha number $B_{2d+1}$; indeed, one could derive the form (\ref{cd}) by
using this as a starting point.  In this approach, one would start with the
property that $c^{(d)}$ has the factor
$(q-B_{2d+1})=(q-q_{d,1})$.  Since $B_{2d+1}$ is, in general, irrational, it is
necessary to symmetrize the product $c^{(d)}=\prod (q-q_z)$ over the zeros
$q_z$ in order to obtain rational (indeed, integer) coefficients for the
$c^{(d)}$.  This symmetrization yields (\ref{cdprod}).  To see this, we note
that to incorporate this zero of $c^{(d)}$ at $q=B_{2d+1}$ one can start with
the identity $[\exp(2\pi i/(2d+1))]^{2d+1}-1=0$, express this as
$[\cos(\frac{2\pi}{2d+1})+i\sin(\frac{2\pi}{2d+1})]^{2d+1}$, expand the
resulting expression, and then impose the condition that the polynomial
vanishes at $q=q_{d,1}=B_{2d+1}$ by setting $\cos(2\pi/(2d+1))=q/2 - 1$.  This
actually yields two equations, one for the real part and one for the imaginary
part. Concentrating on the real part, we observe that the resultant equation 
has the form $(1/2)(q-4)(c^{(d)})^2$.  This yields the result 
\beq 
c^{(d)} = \biggl [ \frac{2}{q-4}\Bigl ( T_{2d+1}(\frac{q}{2}-1)-1 \Bigr ) 
\biggr ]^{1/2}
\label{cdcheb}
\eeq
where $T_n(x)$ is the Chebyshev polynomial of the first kind, defined by 
(e.g., \cite{gr,riordan}) 
\beqs
T_n(x) & = & Re \biggl [ (x+i\sqrt{1-x^2} \ )^n \biggr ] \cr\cr
       & = & \frac{1}{2}\sum_{j=0}^{[\frac{n}{2}]} (-1)^j \frac{n}{n-j}
{n-j \choose j} (2x)^{n-2j} 
\label{tndef}
\eeqs
As with $U_n(x)$, in the first line of (\ref{tndef}), $x$ must be real, but
the polynomial in the second line of (\ref{tndef}) constitutes a
definition of $T_n(x)$ for complex $x$; note also that the right-hand side of
the second line is defined to be 1 for $n=0$. 
Simplifying eq. (\ref{cdcheb}), one finally obtains eq. (\ref{cd}) in terms of 
the Chebyshev polynomial of the second kind.  The starting
point is equivalent to the identity $[\exp(2\pi k/(2d+1))]^{2d+1}-1=0$, and
hence it follows that $c^{(d)}$ vanishes at $q_{d,k}$ for 
$2 \le k \le d$ as well as for $k=1$.  This implies the formula (\ref{cdprod}).

\vspace{4mm}

The general formula ({\ref{cd}) or the equivalent result (\ref{cdprod}) gives a
deeper understanding of, the known exact solutions for chromatic polynomials
and the full Potts model partition functions on cyclic strips of the square and
triangular lattice.  For reference, these include the degenerate case $L_y=1$
(circuit graph), and the cases $L_y=2$ in \cite{bds}, $L_y=3$ in \cite{wcy},
and $L_y=4$ in \cite{s4} for cyclic strips of the square lattice; and the cases
$L_y=2$ in \cite{wcy} and $L_y=3$ and $L_y=4$ in \cite{t} for cyclic strips of
the triangular lattice.  For the full Potts model partition functions, the
exact solutions include the elementary $L_y=1$ solution for the circuit graph,
and the $L_y=2$ solutions in \cite{a} and \cite{ta} for cyclic strips of the
square and triangular lattices, respectively.  Explicit examples will be given
below. As noted, the coefficients with degree $d$ occur for $0 \le d \le L_y$.
In the following we shall work out implications of the formula (\ref{cd}) for
the structure of the chromatic polynomial and full Potts model partition
function for arbitrarily wide cyclic strips of the square and triangular
lattices.

 From the formula (\ref{cd}) or (\ref{cdprod}) and properties of the
Chebyshev polynomials of the second kind, one can establish a number of
useful properties of the coefficients $c^{(d)}$.  First, we recall that a 
generating function for the $U_n(x)$ is (e.g. \cite{gr}) 
\beq
\frac{1}{1-2xz + z^2} = \sum_{n=0}^\infty U_n(x)z^n
\label{ungenfun}
\eeq
The formal sum in
(\ref{ungenfun}) converges if $|x| < 1$ and $|z| < 1$; however, for our
purposes, we shall only use it as a means of extracting the coefficient of the
$z^n$ term as $U_n(x)$, and this can be done for any $x$ and $z$. 
We infer the following properties of the $c^{(d)}$:

\begin{enumerate}

\item 

$c^{(d)}$ is a polynomial of degree $d$ in the variable $q$, with integer
coefficients, highest-order term equal to $q^d$, and alternating signs for the
subsequent terms of descending degree in $q$.

\item

 We recall the definition of a unimodal polynomial as one with the property
 that the magnitudes of its coefficients increase monotonically up to a point
 and then (with possible equality of the maximal two coefficients) decrease
 monotonically.  From the property that the magnitudes of the coefficients of
 $U_n(x)$ are unimodal, it follows directly that the same is true of the
 magnitudes of the coefficients of $c^{(d)}$.

\item

The zeros of $c^{(d)}$ are real, positive, simple, lie in the interval 
$0 < q < 4$ and are given as follows:
\beq
c^{(d)}=0 \quad {\rm at} \quad q=q_{d,k} \ , \quad k=1,2,...,d
\label{cdzero}
\eeq
In the limit $d \to \infty$, these zeros become dense in this interval.

\item

If $d \ge 1$, then one of the zeros of $c^{(d)}$ is at a Tutte-Beraha number, 
namely, 
\beq
q_{d,1}=B_{2d+1}
\label{qd1}
\eeq
where the Tutte-Beraha number $B_r$ is defined as \cite{tutte3},\cite{bkw}
\beq
B_r = 2 + 2\cos \Bigl ( \frac{2\pi}{r} \Bigr ) = 4\cos^2 \Bigl (
\frac{\pi}{r} \Bigr ) \ , r=1,2,...
\label{br}
\eeq
The first Tutte-Beraha numbers are $B_1=4$, $B_2=0$, $B_3=1$,
$B_4=2$, $B_5=(3+\sqrt{5} \ )/2=2.618..$, $B_6=3$. 

\item 

Depending on the value of $d$, $c^{(d)}$ may have other zero(s) 
at Tutte-Beraha number(s) distinct from $B_{2d+1}$ in (\ref{qd1}). 
This cannot happen if $2d+1$ is prime. 
If $2d+1$ is not prime, consider the case where there exists $d^\prime$ such
that $k(2d^\prime+1)=2d+1$.  Since $2d+1$ and $2d^\prime+1$ are both odd, it
follows that $k$ is also odd, so we can write this equation in a symmetric
manner as 
\beq 
(2d^\prime+1)(2d^{\prime \prime}+1)=2d+1 \ .
\label{dpdpp}
\eeq 
Then $c^{(d)}$ has additional zero(s) at $B_{2d^\prime+1}$ and
$B_{2d^{\prime\prime}+1}$.  These degenerate into a single additional zero if
$2d+1=p^2$ where $p$ is prime.  Note that the factorization (\ref{dpdpp}) is
not unique.  For example, consider $c^{(d)}$ for $d=52$.  Then $2d+1$ has the
full and unique factorization $105=3 \cdot 5 \cdot 7$.  There are thus three
ways of writing the twofold factorization in (\ref{dpdpp}): $3 \cdot 35$, $5
\cdot 21$, and $7 \cdot 15$, and hence, in addition to
the zero at the Tutte-Beraha number $B_{2d+1}=B_{105}$, the coefficient 
$c^{(105)}$ also has
zeros at the six other Tutte-Beraha numbers $B_3$, $B_5$, $B_7$, $B_{15}$,
$B_{21}$, and $B_{35}$.  As this example makes clear, in other cases there
could be more than three twofold factorizations of the form (\ref{dpdpp}). 

\item

As a particular case of the previous item, if $d=1$ mod 3, then, setting
$d=3j+1$ and substituting in the above equation, we find
$(2d^\prime+1)(2d^{\prime \prime}+1)=3(2j+1)$.  If $j=0$ so that $d=1$, i.e.,
$c^{(d)}=q-1$, then there is only the single zero at $B_3=1$ given by
(\ref{qd1}).  If $j=1$, i.e., $d=4$, then $c^{(4)}$ has a zero at $B_3$ and at
$B_9$.  If $j \ge 2$, then in addition to the zero at $B_{2d+1}$ given by
(\ref{qd1}), $c^{(d)}$ has zeros at the Tutte-Beraha numbers $B_3$ and
$B_{2j+1}$ (as well as possible others).  Note that since $c^{(d)}$ has only
simple zeros, it follows that if $d=1$ mod 3, then has a simple factor $(q-1)$.
(The expression given above for $c^{(4)}$ illustrates this.)

\item 

Since $U_{2n}(0)=(-1)^n$, it follows that 
\beq
c^{(d)}=(-1)^d \quad {\rm for} \quad q=0 \ . 
\label{cdq0}
\eeq

\item 

\beq
{\rm If} \quad q=1 \quad {\rm then} \quad 
c^{(d)}= \cases{ 1 & if $d=0$ \ mod \ 3 \cr
                 0 & if $d=1$ \ mod \ 3 \cr 
                -1 & if $d=2$ \ mod \ 3  }
\label{cdq1}
\eeq

\item

\beq
{\rm If} \quad q=2 \quad {\rm then} \quad
c^{(d)}= \cases{ 1 & if $d=0,1$ \ mod \ 4 \cr
                 -1 & if $d=2,3$ \ mod \ 4 \cr } 
\label{cdq2}
\eeq

\item

\beq
{\rm If} \quad q=3 \quad {\rm then} \quad
c^{(d)}= \cases{ 1 & if $d=0,2$ \ mod \ 6 \cr
                 2 & if $d=1$ \ mod \ 6 \cr 
               -1  & if $d=3,5$ \ mod \ 6 \cr
               -2  & if $d=4$ \ mod \ 6 \cr }
\label{cdq3}
\eeq

\item 

\beq
c^{(d)}= 2d+1 \quad {\rm for} \quad q=4 \ . 
\label{cdq4}
\eeq

\item

As a consequence of the inequality $|U_n(x)| \le n+1$ for $-1 \le x \le 1$, we
have the inequality 
\beq
|c^{(d)}| \le 2d+1 \quad {\rm for} \ \ 0 \le q \le 4 \ . 
\label{cdineq}
\eeq
This inequality is saturated at $q=4$, i.e. $x=1$. 

\item 

Using the recursion relation for Chebyshev polynomials, 
\beq
U_{n+1}(x)=2xU_n(x)-U_{n-1}(x)
\label{urecursion}
\eeq
iteratively, we find the resulting recursion for the $c^{(d)}$, 
\beq
c^{(d+1)}=(q-2)c^{(d)}-c^{(d-1)} \ . 
\label{cdrecursion}
\eeq

\end{enumerate}

\vspace{6mm}

If one lets 
\beq
q = 2 + 2 \cos \theta = 4\cos^2 \Bigl ( \frac{\theta}{2} \Bigr ) 
\label{qtrel}
\eeq
one sees that the argument of the Chebyshev polynomial of the second kind in
(\ref{cd}) is given by $x=\frac{\sqrt{q}}{2}=\cos (\theta/2)$.  A relevant 
identity is (with $\omega=\theta/2$ here)
\beq
U_n(\cos \omega) = \frac{\sin((n+1)\omega)}{\sin \omega} \ . 
\label{unsin}
\eeq The fact that the transformation (\ref{qtrel}) applies, with $\theta$
real, for $q \in [0,4]$ shows the special role of this interval for Potts model
partition functions on the strips under consideration here.  This makes an
intriguing connection with the 2D Potts ferromagnet, which has a second-order
phase transition for the values $q=2$ (Ising), $q=3$, and $q=4$; and a
first-order phase transition for $q \ge 5$.  Of course, the physical
thermodynamic behavior of the Potts ferromagnet on the infinite-length, finite
width strips is quite different from that of the model on 2D lattices; in the
former cases, since these are quasi-one-dimensional, the model has only a
zero-temperature critical point rather than a finite-temperature phase
transition.  For the antiferromagnetic Potts model, the transformation means
that the value $q=4$ is special (as the positive value of $q$ where the angle
$\theta$ changes from being real to imaginary) and is in accord with the
special role of $q=4$ in this model on the infinite two-dimensional lattice
\cite{baxter} (see also \cite{baxkwu}).  This is somewhat similar to earlier
situations in which studies of lower-dimensional realizations of spin models 
gave insight into properties of higher-dimensional realizations
(e.g. \cite{mp}-\cite{bls} for $O(N)$ models and \cite{a,ta,is1d} for
Potts/Ising models).  However, in assessing the connection of the special 
role of the value $q=4$ in the coefficients $c^{(d)}$ with the $T=0$ critical 
point at $q=4$ for the Potts antiferromagnet on the triangular lattice, one
must also taken into account the fact that the formula (\ref{cd}) also applies
for the cyclic strips of the square lattice that have been studied; however,
again taking the limit $L_y \to \infty$, the Potts antiferromagnet on the
square lattice has a zero-temperature critical point at $q=3$ rather than 
at $q=4$ \cite{lieb}. 

\vspace{6mm}

One of the interesting aspects of the zeros $q_{d,k}$ of $c^{(d)}$ is that at
these values of $q$, the critical exponents of the 2D Potts model are
rational.  We recall that the thermal, magnetic, and tricritical exponents for 
the paramagnetic to ferromagnetic transition in the 2D $q$-state Potts
ferromagnet are, for $q \le 4$, \cite{dennijs}-\cite{pearson80},\cite{wurev}
\beq
y_t = \frac{3(1-u)}{2-u}
\label{yt}
\eeq

\beq
y_h, \ y_{h,tric.} = \frac{(3-u)(5-u)}{4(2-u)}
\label{yh}
\eeq
where
\beq
u = \frac{2}{\pi}{\rm arccos} \Bigl (\frac{\sqrt{q}}{2} \Bigr ) 
\label{udef}
\eeq
so that $0 \le u \le 1$ for $4 \ge q \ge 0$.  The angle $\theta$ in 
(\ref{qtrel}) is equal to $\pi u$ in (\ref{udef}), and hence 
\beq
q=q_{d,k} \ \Leftrightarrow \ u = \frac{2k}{2d+1} \ . 
\label{qurel}
\eeq
Note that the converse does not hold; that is, there are rational values of $u$
that are not of the form (\ref{qurel}) and hence do not correspond to any of 
the $q_{d,k}$.  An example is $u=1/2$. 

\vspace{8mm}

\section{Determination of $n_P(L_y,d)$ for Cyclic Strips of the
Square and Triangular Lattices}

In this section and the next we use (\ref{cd}) together with two theorems
(eqs. (\ref{ncrelcyc} and (\ref{ntcrelcyc}) below) to determine structural
properties of the chromatic polynomial and the full Potts model partition
function for cyclic strip graphs $G$ of the square and triangular lattices.
Let us define $n_P(L_y,d)$ as the number of terms $\lambda_{P,G,j}$ in $P(G,q)$
that have as their coefficients $c_{P,G,j}=c^{(d)}$ and $n_Z(L_y,d)$ as the
number of terms $\lambda_{Z,G,j}$ in $Z(G,q,v)$ that have as their
coefficients $c_{Z,G,j}=c^{(d)}$.  For the cyclic strip graphs under
consideration here, these coefficients are independent of $L_x$ and depend on
$L_y$ and $d$; furthermore, they are the same for both square and triangular
strips, as discussed further below. While the individual $\lambda_{P,G,j}$'s in
(\ref{pgsum}) are, in general, different for the cyclic strips of the square
and triangular lattices, the total number of $\lambda_{P,G,j}$'s is the same,
so we use the short notation $N_{P,L_y,\lambda}$.  The same is true for the
full partition function, so we use the notation $N_{Z,L_y,\lambda}$.

Since the chromatic polynomial $P(G,q)$ is the 
special case $v=-1$ (i.e. zero-temperature antiferromagnet) of the 
general (finite-temperature, $J$ positive or negative) Potts model partition 
function, (\ref{zp}), it follows that 
\beq
n_P(L_y,d) \le n_Z(L_y,d) \ . 
\label{npnz}
\eeq
The total number, $N_{P,G,\lambda}$, of different terms $\lambda_{P,G,j}$ in 
(\ref{pgsum}) is given by 
\beq
N_{P,L_y,\lambda} = \sum_{d=0}^{L_y} n_P(L_y,d) \ . 
\label{npsum}
\eeq
Since each term $\lambda_{P,G,j}$ is a distinct eigenvalue of the coloring
matrix ${\cal T}$, the number $N_{P,L_y,\lambda}$ is the number of different 
invariant subspaces in the full ${\cal N}$--dimensional space of coloring
configurations of the transverse slices of the strips. 
Similarly, for the full Potts model partition function, 
\beq
N_{Z,L_y,\lambda} = \sum_{d=0}^{L_y} n_Z(L_y,d) \ . 
\label{nzsum}
\eeq
and analogously this represents the number of different invariant subspaces for
the matrix ${\cal T}_Z$. 
It was shown in \cite{pm} that the $\lambda_{P,G(L_y),j}$'s are the same for 
the cyclic and M\"obius strips (although the corresponding $c_{P,G,j}$'s are 
different)
\beq
\lambda_{P,G(L_y),FBC_y,PBC_x,j}=\lambda_{P,G(L_y),FBC_y,TPBC_x,j} \quad
\forall \ j
\label{lameqcycmb}
\eeq
and hence the total number of terms is also the same:
\beq
N_{P,G(L_y),FBC_y,PBC_x,\lambda}=N_{P,G(L_y),FBC_y,TPBC_x,\lambda} \ . 
\label{ncycmb}
\eeq
The argument in \cite{pm} relied upon the local nature of the
deletion-contraction operations used in calculating $P(G,q)$, and the same
property is true of the deletion-contraction theorem used for calculating the
Tutte polynomial, or equivalently, the Potts model partition function, so that
the following generalizations of (\ref{ncycmb}) holds \cite{bcc,a}:
\beq
\lambda_{Z,G(L_y),FBC_y,PBC_x,j}=\lambda_{Z,G(L_y),FBC_y,TPBC_x,j} \quad
\forall \ j
\label{lamzeqcycmb}
\eeq
and hence 
\beq
N_{Z,G(L_y),FBC_y,PBC_x,\lambda}=N_{Z,G(L_y),FBC_y,TPBC_x,\lambda} \ . 
\label{nzcycmb}
\eeq
For the sum of the coefficients in (\ref{pgsum}), i.e.,
(\ref{cpsum}), we have 
\beq
C_{P,L_y} = \sum_{j=1}^{N_{P,L_y,\lambda}} c_{P,L_y,j} = 
\sum_{d=0}^{L_y} n_P(L_y,d)c^{(d)} 
\label{cdsum}
\eeq
and for the corresponding sum of coefficients in (\ref{zgsum}), 
\beq
C_{Z,L_y} = \sum_{j=1}^{N_{Z,L_y,\lambda}} c_{Z,L_y,j} =
 \sum_{d=0}^{L_y} n_Z(L_y,d)c^{(d)} \ . 
\label{cdsumz}
\eeq

We first recall a theorem specifying $C_{P,L_y}$ for cyclic strips of the
square and triangular lattice \cite{pm}: 

\vspace{8mm}

\begin{flushleft}

Theorem 1. \ 
\beq
C_{P,L_y} = P(T_{L_y},q)=q(q-1)^{L_y-1} \ . 
\label{ncrelcyc}
\eeq

\vspace{8mm}

Proof. \ Using coloring matrix methods \cite{b,matmeth}, one has that 
$C_{P,L_y}$ is equal to the chromatic polynomial for the coloring of the
transverse slice of the strip, which is a line with $L_y$ vertices.  This is a
special case of a tree graph, for which the elementary general result in 
eq. (\ref{pt}) holds. \ $\Box$ 

\end{flushleft}

Next, we have 

\vspace{8mm}

\begin{flushleft}

Theorem 2.  The $n_P(L_y,d)$, $d=0,1,..L_y$ are determined as follows. One has
\beq
n_P(L_y,d)=0 \quad {\rm for} \quad d > L_y \ , 
\label{npup}
\eeq
\beq
n_P(L_y,L_y)=1
\label{npcly}
\eeq
and
\beq
n_P(1,0)=1
\label{np10}
\eeq
with all other numbers $n_P(L_y,d)$ being determined by the two recursion
relations
\beq
n_P(L_y+1,0)=n_P(L_y,1)
\label{nprecursion1}
\eeq
and
\beq
n_P(L_y+1,d) = n_P(L_y,d-1)+n_P(L_y,d)+n_P(L_y,d+1) 
\quad {\rm for} \quad L_y \ge 1 \quad {\rm and} \quad 1 \le d \le L_y+1 \ . 
\label{nprecursion2}
\eeq

\vspace{4mm}

Proof. \  We substitute for $c^{(d)}$ from eq. (\ref{cd}) in eq.
(\ref{ncrelcyc}).  We obtain another equation by differentiating this with
respect to $q$ once; another by differentiating twice, and so forth up to 
$L_y$-fold differentiations.  This yields $L_y+1$ linear equations in the 
$L_y+1$ unknowns, $n_P(L_y,d)$, $d=0,1,...,L_y$.  We solve this set of
equations to get the $n_P(L_y,d)$.  $\Box$ 

\vspace{3mm}

A corollary is that 
\beq
n_P(L_y,L_y-1)=L_y \ . 
\label{npclyminus1}
\eeq

\end{flushleft}

The numbers $n_P(L_y,d)$ can be viewed as integer sequences in $L_y$ for a
given value of $d$.  We find that (for $L_y \ge 1$ where our strips are 
defined) $n_P(L_y,0)$ is a Motzkin number \cite{motzkin}-\cite{sl}
\beq
n_P(L_y,0)=M_{L_y-1}
\label{nplyd0}
\eeq
where the Motzkin number $M_n$ is given by 
\beq
M_n =  \sum_{j=0}^n (-1)^j C_{n+1-j} {n \choose j}
\label{motzkin}
\eeq
where 
\beq
C_n=\frac{1}{n+1}{2n \choose n}
\label{catalan}
\eeq
is the Catalan number. (We also use the symbol $C_n$ to denote the circuit 
graph with $n$ vertices; the meaning will be clear from context.)  The 
Catalan and Motzkin numbers occur in many combinatoric applications
\cite{motzkin}-\cite{sl}.  Among these is the 
construction of non-intersecting chords on a circle; the number of ways of
connecting a subset of $n$ points on a circle by non-intersecting chords is
$M_n$, while the number of ways of completely connecting $2n$ points on the
circle by such chords is $C_n$.  Summing over subsets of points connected by
chords, this yields the well-known relation
\beq
M_n=\sum_{k=0}^{[\frac{n}{2}]} {n \choose 2k} C_k \ . 
\label{cattomot}
\eeq
As a consequence of our relation (\ref{nprecursion1}), the equation 
(\ref{nplyd0}) also implies that
\beq
n_P(L_y,1)=M_{L_y} \ . 
\label{nply1m}
\eeq
A generating function is 
\beq
G_{n_P(L_y,0)}(x) = \frac{1-x-\sqrt{1-2x-3x^2}}{2x^2} = \sum_{L_y=1}^\infty
n_P(L_y,0)x^{L_y-1} \ . 
\label{gpx}
\eeq 
In eq. (\ref{nptotgenfun}) below we shall give an exact determination of
the total number, $N_{P,L_y,\lambda}$, of types of terms $\lambda_{P,L_y,j}$.
In Table \ref{npctablecyc} we list the numbers $n_P(L_y,d)$ and the sums
$N_{P,L_y,\lambda}$ for the first several widths of cyclic strips of the square
and triangular lattice, $1 \le L_y \le 10$.  Others can easily be calculated
using our general formulas.

\begin{table}
\caption{\footnotesize{Table of numbers $n_P(L_y,d)$ and their sums, 
$N_{P,L_y,\lambda}$ for cyclic strips of the square and triangular lattices. 
Blank entries are zero.}}
\begin{center}
\begin{tabular}{|c|c|c|c|c|c|c|c|c|c|c|c|c|}
\hline\hline 
$L_y \ \downarrow$ \ \ $d \ \rightarrow$ 
   & 0 & 1   & 2   & 3   & 4   & 5  & 6  & 7 & 8 & 9 & 10 & 
$N_{P,L_y,\lambda}$ 
\\ \hline\hline
1  & 1   & 1   &     &     &     &    &    &   &   &    &   & 2    \\ \hline
2  & 1   & 2   & 1   &     &     &    &    &   &   &    &   & 4    \\ \hline
3  & 2   & 4   & 3   & 1   &     &    &    &   &   &    &   & 10   \\ \hline 
4  & 4   & 9   & 8   & 4   & 1   &    &    &   &   &    &   & 26   \\ \hline
5  & 9   & 21  & 21  & 13  & 5   & 1  &    &   &   &    &   & 70   \\ \hline
6  & 21  & 51  & 55  & 39  & 19  & 6  & 1  &   &   &    &   & 192  \\ \hline
7  & 51  & 127 & 145 & 113 & 64  & 26 & 7  & 1 &   &    &   & 534  \\ \hline
8  & 127 & 323 & 385 & 322 & 203 & 97 & 34 & 8 & 1 &    &   & 1500 \\ \hline
9  & 323 & 835 & 1030& 910 & 622 & 334& 139& 43& 9 & 1  &   & 4246 \\ \hline
10 & 835 & 2188&2775 &2562 &1866 &1095& 516&191& 53& 10 & 1 & 12092 \\ 
\hline\hline  
\end{tabular}
\end{center}
\label{npctablecyc}
\end{table}

Now let us consider a random walk on the nonnegative integers such that in each
step the walker moves by $+1$, $-1$, or 0 units.  Denoting $m(n,k)$ as the 
number of walks of length $n$ steps starting at $0$ and ending at $k$, we 
obtain the Motzkin triangle 
in Table \ref{motzkintri}. The first column, corresponding to $k=0$ is the
number of walks defined above that return to the origin after $n$ steps. This
is given by the Motzkin number defined in (\ref{motzkin}); $m(n,0)=M_n$.  
The second column, $m(n,1)$, is given by the first
differences of Motzkin numbers: 
\beq
m(n,1) =  \sum_{j=0}^{[\frac{n+1}{2}]} \frac{1}{j+1} {n \choose 2j-1} 
{2j \choose j} \ .
\label{motzkindiff}
\eeq
The row sums in Table \ref{motzkintri} will be important below; we denote them
as 
\beq
{\cal S}_n = \sum_{k=0}^n  m(n,k) \ .
\label{rowsum}
\eeq
Notice that Table \ref{npctablecyc} can be viewed as the combination of two
Motzkin triangles, i.e. Table \ref{motzkintri}, as follows,
\beq
n_P(L_y,d) =  m(L_y-1,d) + m(L_y-1,d-1) \ .
\label{nplydm}
\eeq

\begin{table}
\caption{\footnotesize{Table of numbers $m(n,k)$ for the number of  a
random walk from $0$ to $k$ in $n$ steps and the row sums ${\cal S}_n$. 
Blank entries are zero.}}
\begin{center}
\begin{tabular}{|c|c|c|c|c|c|c|c|c|c|c|c|}
\hline\hline 
$n \ \downarrow$ \ \ $k \ \rightarrow$ 
   & 0 & 1   & 2   & 3   & 4   & 5  & 6  & 7 & 8 & 9 & ${\cal S}_n$ 
\\ \hline\hline
0  & 1   &     &     &     &     &    &    &   &   &    & 1    \\ \hline
1  & 1   & 1   &     &     &     &    &    &   &   &    & 2    \\ \hline
2  & 2   & 2   & 1   &     &     &    &    &   &   &    & 5    \\ \hline
3  & 4   & 5   & 3   & 1   &     &    &    &   &   &    & 13   \\ \hline 
4  & 9   & 12  & 9   & 4   & 1   &    &    &   &   &    & 35   \\ \hline
5  & 21  & 30  & 25  & 14  & 5   & 1  &    &   &   &    & 96   \\ \hline
6  & 51  & 76  & 69  & 44  & 20  & 6  & 1  &   &   &    & 267  \\ \hline
7  & 127 & 196 & 189 & 133 & 70  & 27 & 7  & 1 &   &    & 750  \\ \hline
8  & 323 & 512 & 518 & 392 & 230 & 104& 35 & 8 & 1 &    & 2123 \\ \hline
9  & 835 & 1353&1422 &1140 & 726 & 369& 147& 44& 9 & 1  & 4246 \\ \hline\hline
\end{tabular}
\end{center}
\label{motzkintri}
\end{table}

In light of the property that the coefficients (= multiplicities of eigenvalues
of the transfer matrix) 
$c^{(d)}$ are Chebyshev polynomials of the second kind for the cyclic strips of
the square and triangular lattices and for M\"obius strips of the square
lattice (discussed below), and the finding in eq. (\ref{nplyd0}) for the
numbers of $\lambda_{P,G,j}$'s with coefficients $c^{(0)}$ and $c^{(1)}$, it is
interesting to note that in \cite{aigner} a connection has been shown between
what is called a Motzkin polynomial and this Chebyshev polynomial.
To see this, one defines $s_{h,0}=M_h$, $s_{h,1}=M_h-M_{h-1}$ and shows that,
with $s_{h,n}=s_{h,n-1}-s_{h-1,n-1}-s_{h-2,n-2}$ for $1 \le n \le h$, it 
is possible to express $s_{h,n}$ as $s_{h,n}=a_n M_h + a_{n-1}M_{h-1} +
... + a_0M_{h-n}$, where the coefficients $a_j$ are independent of $h$.
The Motzkin polynomial is defined as $S_n(x)=a_nx^n + ... + a_1x + a_0$,
and the connection is that $S_n(x)=U_n(\frac{x-1}{2})$ \cite{aigner}.

Certain ($q$-independent) relations between the $n_P(L_y,d)$ can
be derived by evaluating the sum (\ref{cdsum}) and its result (\ref{ncrelcyc})
for $q=0,1$, and 2. Setting $q=0$ in (\ref{cdsum}) and (\ref{ncrelcyc}), and 
using (\ref{cdq0}), we have 
\beq
\sum_{d=0}^{L_y} n_P(L_y,d)c^{(d)}(q=0)=
\sum_{d=0}^{L_y} (-1)^d n_P(L_y,d)=0 \ .
\label{cdsumq0}
\eeq
Next we evaluate (\ref{cdsum}) and (\ref{ncrelcyc}) for $q=1$. If $L_y=1$, 
this only involves one term, $n_P(1,0)=1$.  So consider $L_y \ge 2$ and set 
$L_y=3k+r$ with integer $k \ge 0$ and $r=0,1$, or 2.  
To express the equation compactly, it is 
convenient to use a Heaviside step function $\theta(z)$, where $\theta(z)=1$ if
$z > 0$ and $\theta(z)=0$ if $z \le 0$.  Using (\ref{cdq1}), we have
\beqs
\sum_{d=0}^{L_y} n_P(L_y,d)c^{(d)}(q=1) & = & \sum_{j=0}^k n_P(L_y,3j) -
\sum_{j=0}^{k-1+\theta(r-\frac{3}{2})} n_P(L_y,3j+2) \cr\cr
& = & 0 
\label{cdsumq1}
\eeqs
where here and in other summations, it is understood that if the upper limit on
the summation is negative, the sum is zero.  A similar relation can be obtained
in a straightforward manner by substituting $q=2$ in (\ref{cdsum}) and 
(\ref{ncrelcyc}). Other similar relations among the $n_P(L_y,d)$ 
can be obtained by evaluating (\ref{cdsum}) and (\ref{ncrelcyc}) for $q=3$ and
$q=4$; for these relations, the right-hand sides grow with $L_y$, in contrast
to (\ref{cdsumq0})-(\ref{cdsumq1}), where the right-hand sides are constant.

\section{Determination of $n_Z(L_y,d)$ for Cyclic Strips of the
Square and Triangular Lattices}

We next generalize these results to the full Potts model partition function.
We determine the numbers $n_Z(L_y,d)$ by observing that in the cyclic strips
under consideration here, the only difference between the coloring for the
$T=0$ Potts antiferromagnet and for the general Potts model at finite
temperature is the relaxation of the constraint that no two adjacent vertices
can have the same color (indeed, for ferromagnetic coupling, there is a
preference for these vertices to have the same color).  Hence, one simply
counts all possible colorings, and obtains

\begin{flushleft}

Theorem 3. 
\beq
\sum_{d=0}^{L_y} n_Z(L_y,d)c^{(d)} = q^{L_y} \ . 
\label{ntcrelcyc}
\eeq
Next, we have

\vspace{6mm}

Theorem 4.  The $n_Z(L_y,d)$ are determined as follows. One has
\beq
n_Z(L_y,d)=0 \quad {\rm for} \quad d > L_y \ , 
\label{ntup}
\eeq
\beq
n_Z(L_y,L_y)=1
\label{ntcly}
\eeq
and
\beq
n_Z(1,0)=1 \ . 
\label{ntc10}
\eeq
All other numbers $n_Z(L_y,d)$ are then determined by the two recursion
relations
\beq
n_Z(L_y+1,0) = n_Z(L_y,0) + n_Z(L_y,1)
\label{ntrecursion1}
\eeq
and
\beqs
n_Z(L_y+1,d) & = & n_Z(L_y,d-1) + 2n_Z(L_y,d) + n_Z(L_y,d+1) \cr\cr
& & \quad {\rm for} \quad 1 \le d \le L_y+1 \ . 
\label{ntrecursion2}
\eeqs

\vspace{5mm}

Proof. \ The proof is similar to that for Theorem 2: we substitute for
$c^{(d)}$ from eq. (\ref{cd}) in eq.  (\ref{ntcrelcyc}).  We obtain another
equation by differentiating this with respect to $q$ once; another by
differentiating twice, and so forth up to $L_y$-fold differentiations.  This
yields $L_y+1$ linear equations in the $L_y+1$ unknowns, $n_Z(L_y,d)$,
$d=0,1,...,L_y$.  We solve this set of equations to get the $n_Z(L_y,d)$.
$\Box$

A corollary is that 
\beq
n_Z(L_y,L_y-1)=2L_y-1 \ . 
\label{ntclyminus1}
\eeq

\end{flushleft}

 From Theorems 3 and 4 we find a general formula for the numbers 
$n_Z(L_y,d)$:
\beq
n_Z(L_y,d)=\frac{(2d+1)}{(L_y+d+1)}{2L_y \choose L_y-d}
\label{nzlyd}
\eeq
for $0 \le d \le L_y$, with $n_Z(L_y,d)=0$ for $d > L_y$. 
For fixed $d$ in the range $0 \le d \le L_y$ where $n_Z(L_y,d)$ is
nonvanishing, it has the leading asymptotic behavior
\beq
n_Z(L_y,d) \sim (2d+1)\pi^{-1/2} L_y^{-3/2} \ 4^{L_y} \Bigl [ 1 + O(L_y^{-1})
\Bigr ] \quad {\rm as} \ \ L_y \to \infty \ \ {\rm for \ fixed} \ \ d \ . 
\label{nzlydasymp}
\eeq
(The formal notation $1+O(L_y^{-1})$ in (\ref{nzlydasymp}) and in other 
asymptotic expansions 
is not intended to indicate the sign of the coefficient of the $O(L_y^{-1})$ 
term; here it is actually negative.) 
As a measure of the asymptotic behavior of $n_Z(L_y,d)$ when $d$, rather than
being fixed, is a finite fraction of $L_y$, we take the central value 
$d=L_y/2$ and calculate that 
\beq
n_Z(L_y,\frac{L_y}{2}) \sim \pi^{-1/2} L_y^{-1/2} 4^{2L_y+1} 
3^{-\frac{3}{2}(L_y+1)}\Bigl [ 1 + O(L_y^{-1}) \Bigr ]
 \quad {\rm as} \ \ L_y \to \infty \ . 
\label{nzlydmiddleasymp}
\eeq
Note that for the special case $d=0$, this reduces to 
\beq
n_Z(L_y,0)=C_{L_y}
\label{nzlyd0}
\eeq
where the Catalan number $C_n$ was defined in (\ref{catalan}).  
In Table \ref{ntctablecyc} we list the first few numbers 
$n_Z(L_y,d)$ and the total sums $N_{Z,L_y,\lambda}$ that will be calculated
below in (\ref{nztot}). 

\begin{table}
\caption{\footnotesize{Table of numbers $n_Z(L_y,d)$ and their sums, 
$N_{Z,G,\lambda}$.  See text for general formulas.}}
\begin{center}
\begin{tabular}{|c|c|c|c|c|c|c|c|c|c|c|c|c|}
\hline\hline 
$L_y \ \downarrow$ \ \ $d \ \rightarrow$ 
   & 0 & 1   & 2   & 3   & 4   & 5  & 6  & 7 & 8 & 9 & 10 & 
$N_{Z,L_y,\lambda}$ 
\\ \hline\hline
1  & 1   & 1   &     &     &     &    &    &   &   &   &   & 2     \\ \hline
2  & 2   & 3   & 1   &     &     &    &    &   &   &   &   & 6     \\ \hline
3  & 5   & 9   & 5   & 1   &     &    &    &   &   &   &   & 20    \\ \hline
4  & 14  & 28  & 20  & 7   & 1   &    &    &   &   &   &   & 70    \\ \hline
5  & 42  & 90  & 75  & 35  & 9   & 1  &    &   &   &   &   & 252   \\ \hline
6  & 132 & 297 & 275 & 154 & 54  & 11 & 1  &   &   &   &   & 924   \\ \hline
7  & 429 & 1001& 1001& 637 & 273 & 77 & 13 & 1 &   &   &   & 3432  \\ \hline
8  & 1430& 3432& 3640& 2548& 1260& 440& 104& 15& 1 &   &   & 12870 \\ \hline
9  & 4862&11934&13260& 9996& 5508&2244& 663&135& 17& 1 &   & 48620 \\ \hline
10 &16796&41990&48450&38760&23256&10659&3705&950&170&19& 1 & 184756 \\ 
\hline\hline  
\end{tabular}
\end{center}
\label{ntctablecyc}
\end{table}

Similar to Table \ref{motzkintri}, we can also make up a Catalan triangle, and
shall find that Table \ref{ntctablecyc} is just the combination of two
triangles.

Certain ($q$-independent) relations between the $n_Z(L_y,d)$ can be derived 
by evaluating the sum (\ref{cdsumz}) and its result (\ref{ntcrelcyc}) for 
$q=0$ and $q=1$. Setting $q=0$ in this sums and using (\ref{cdq0}), we have 
\beq
\sum_{d=0}^{L_y} (-1)^d n_Z(L_y,d)=0 \ .
\label{ctdsumq0}
\eeq
Next we evaluate (\ref{cdsumz}) and (\ref{ntcrelcyc}) for $q=1$.  
Let $L_y=3k+r$ with $k\ge 0$ and $r=0,1$, or 2.  Using (\ref{cdq1}), we have
\beq
\sum_{j=0}^k n_Z(L_y,3j) - \sum_{j=0}^{k-1+\theta(r-\frac{3}{2})} 
n_Z(L_y,3j+2) = 1 \ .
\label{ctdsumq1}
\eeq

\vspace{6mm}

We have obtained the following relations involving both $n_Z(L_y,d)$ and 
$n_P(L_y,d)$, as well as the total numbers $N_{Z,L_y,\lambda}$ and 
$N_{P,L_y,\lambda}$ 

\beq
n_P(L_y,d)=\sum_{j=0}^{L_y-1}(-1)^j {L_y-1 \choose j}n_Z(L_y-j,d)
\label{npnt}
\eeq
and
\beq
n_Z(L_y,d) = \sum_{j=0}^{L_y-1} {L_y-1 \choose j}n_P(L_y-j,d)  \ . 
\label{ntnpsum}
\eeq
Since the total numbers of terms $N_{P,L_y,\lambda}$ and $N_{Z,L_y,\lambda}$
are sums of the $n_P(L_y,d)$ and $n_Z(L_y,d)$ given, respectively, by 
(\ref{npsum}) and (\ref{nzsum}), it follows that 
\beq
N_{P,L_y,\lambda}=\sum_{j=0}^{L_y-1}(-1)^j {L_y-1 \choose j}
N_{Z,L_y-j,\lambda}
\label{nptotnt}
\eeq
and
\beq
N_{Z,L_y,\lambda} = \sum_{j=0}^{L_y-1} {L_y-1 \choose j}
N_{P,L_y-j,\lambda} \ . 
\label{nttotnpsum}
\eeq
Note that for $d=0$, our eq. (\ref{npnt}) reduces to the relation
(\ref{motzkin}) expressing the Motzkin number as a certain weighted sum of
Catalan numbers, while eq. (\ref{ntnpsum}) reduces to the relation expressing 
the Catalan number as a weighted sum of Motzkin numbers (e.g., \cite{bernhart})
\beq
C_n=\sum_{j=0}^{n-1}{n-1 \choose j} M_{n-j-1} = \sum_{k=0}^{n-1}
{n-1 \choose k} M_k 
\label{motztocat}
\eeq
for $n \ge 1$ (with $C_0=1$). 
Further, for the even and odd widths $L_y=2\ell$ and $L_y=2\ell+1$, 
\beq
N_{Z,\ell,\lambda}=
\sum_{j=0}^{2\ell-1}(-1)^j {2\ell-1 \choose j} N_{P,2\ell-j,\lambda}
\label{npeven}
\eeq
and
\beq
2N_{Z,\ell,\lambda}=
\sum_{j=0}^{2\ell} (-1)^j {2\ell \choose j} N_{P,2\ell+1-j,\lambda} \ .
\label{npodd}
\eeq

\vspace{6mm}

\section{Determination of $N_{P,L_y,\lambda}$ and $N_{Z,L_y,\lambda}$ for 
Cyclic and M\"obius Strips of the Square and Triangular Lattices and Relation
with Directed Lattice Animals}

In this section we shall use our results above to calculate the total number,
$N_{P,G,\lambda}$, of $\lambda_{P,G,j}$'s in the chromatic polynomial
(\ref{pgsum}) and the total number, $N_{Z,G,\lambda}$, of $\lambda_{Z,G,j}$'s
in the full Potts model partition function for cyclic strips $G$ of the square
and triangular lattice.  Since the individual numbers $n_P(L_y,d)$ were shown
to be the same for the cyclic strips of the square and triangular lattices, and
similarly for $n_Z(L_y,d)$, clearly it is also true that the respective total
numbers $N_{P,G,\lambda}$ and $N_{Z,G,\lambda}$ are the same for cyclic strips
of the square and triangular lattices. Furthermore, as a consequence of
eqs. (\ref{lameqcycmb})-(\ref{nzcycmb}), our use of cyclic strips to calculate
the total numbers $N_{P,G,\lambda}$ and $N_{Z,G,\lambda}$ also yields these
respective numbers for the M\"obius strips of the square and triangular
lattices.  This is useful since, as will be seen in the next sections, the
individual coefficients involve a larger set for M\"obius strips of the square
lattice, namely not just $c^{(d)}$ but also $-c^{(d)}$; moreover, the
coefficients for M\"obius strips of the triangular lattice are not, in general,
polynomials in $q$ \cite{wcy}.

\subsection{ $N_{P,L_y,\lambda}$ }

For the total number of terms $N_{P,L_y,\lambda}$, from our theorems above we 
obtain the recursion relation 
\beq
N_{P,L_y+1,\lambda}=3N_{P,L_y,\lambda}-2n_P(L_y,0) \ . 
\label{nptotrecursion}
\eeq 
We find that 
\beq
N_{P,L_y,\lambda}=2{\cal S}_{L_y}=
2(L_y-1)! \ \sum_{j=0}^{[\frac{L_y}{2}]} \frac{(L_y-j)}{
(j!)^2(L_y-2j)!}
\label{nptotform}
\eeq
where ${\cal S}_n$ was given in (\ref{rowsum}). 
A generating function for these numbers is 
\beq
G_{N_{P,L_y,\lambda}}(x) = \biggl ( \frac{1+x}{1-3x} \biggr)^{1/2}-1 = 
\sum_{L_y=1}^\infty N_{P,L_y,\lambda} \ x^{L_y} \ .
\label{nptotgenfun}
\eeq 
From this, it follows that the number $N_{P,L_y,\lambda}$ grows
 exponentially fast with the width $L_y$ of the cyclic or M\"obius strip of
 the square or triangular lattice, with the leading asymptotic behavior 
\beq
 N_{P,L_y,\lambda} \sim L_y^{-1/2} \ 3^{L_y} \quad {\rm as} \ \ L_y \to \infty
 \ .
\label{npasymp}
\eeq

\subsection{ $N_{Z,G,\lambda}$}

A corollary of Theorem 4 is that 
\beq
N_{Z,L_y+1,\lambda}=4N_{Z,L_y,\lambda}-2n_Z(L_y,0) \ . 
\label{nttotrecursion}
\eeq

Summing the individual numbers $n_Z(L_y,d)$ to evaluate the total, 
(\ref{nzsum}), we have, for cyclic and M\"obius strips of width $L_y$ of the 
square and triangular lattices, 
\beqs
N_{Z,L_y,\lambda} & = & \sum_{d=0}^{L_y} \frac{(2d+1)}{(L_y+d+1)}
{2L_y \choose L_y-d} \cr\cr 
& = & {2L_y \choose L_y} \ . 
\label{nztot}
\eeqs

As $L_y \to \infty$, $N_{Z,L_y,\lambda}$ has the leading asymptotic behavior 
\beq
N_{Z,L_y,\lambda} \sim \pi^{-1/2}L_y^{-1/2} \ 4^{L_y}\Bigl [ 1 + O(L_y^{-1})
\Bigr ] \quad 
{\rm as} \ \ L_y \to \infty \ . 
\label{nztotasymp}
\eeq
We give two comparisons of the asymptotic behavior of individual numbers 
$n_Z(L_y,d)$ with the total, $N_{Z,L_y,\lambda}$.  First, we compare the
growth of $n_Z(L_y,d)$ for fixed $d$ (in the range $0 \le d \le L_y$
where $n_Z(L_y,d)$ is nonzero) with the growth of $N_{Z,L_y,\lambda}$: 
\beq
\frac{n_Z(L_y,d)}{N_{Z,L_y,\lambda}} \sim \frac{(2d+1)}{L_y}\Bigl [ 1 +
O(L_y^{-1}) \Bigr ] 
\quad {\rm as} \ \ L_y \to \infty \ \ {\rm for \ fixed} \ \ d \ .
\label{nzdntotratio}
\eeq
Second, we compare the relative growths of the central number
$n_Z(L_y,\frac{L_y}{2})$ and the total, $N_{Z,L_y,\lambda}$:
\beqs
\frac{n_Z(L_y,\frac{L_y}{2})}{N_{Z,L_y,\lambda}} & \sim & \biggl ( 
\frac{4}{3^{3/2}} \biggr )^{L_y+1}\biggl [ 1 + O(L_y^{-1}) \biggr ] \cr\cr
& \sim & (0.769800...)^{L_y+1}\biggl [ 1 + O(L_y^{-1}) \biggr ] \cr\cr
& & \quad {\rm as} \ \ L_y \to \infty \ . 
\label{nzdmiddlentotratio}
\eeqs

\subsection{Connection with Directed Lattice Animals}

For an arbitrary $G$, the Potts model partition function $Z(G,q,v)$ gives
information about certain graph-theoretic quantities describing $G$, as is
clear from the Kasteleyn-Fortuin representation (\ref{cluster}) and the
equivalence in eqs. (\ref{ztutte}) and (\ref{zwhit}) to the Tutte polynomial
and Whitney rank polynomial.  An example of this information is illustrated by
the zero-temperature antiferromagnetic special case $P(G,q)=Z(G,q,-1)$ in
eq. (\ref{zp}), i.e., the chromatic polynomial, counting the number of proper
vertex colorings of $G$.  Other examples are that the Tutte polynomials
$T(G,x,y)$ for various special values of its arguments, counts the number
of spanning trees, spanning forests, connected spanning subgraphs, and spanning
subgraphs of $G$ (see appendix).  It is also well known that the $q \to 1$
limit of the Potts model is related to bond percolation and the $q \to 1$ limit
of a certain multisite Potts model is related to site percolation, as reviewed
in \cite{wurev}. 

\vspace{4mm}

Here we would like to report two new and very interesting relations with
graph-theoretic quantities that we have found from our calculations.  
These involve directed lattice animals. 
To explain this, we recall that an animal $A$ on a lattice or
more generally a graph $G$ is defined as a (finite) set of vertices in $G$ with
the property that any two vertices in $A$ are connected by means of a path in
$G$ having all its vertices in $A$. Thus, any vertex in $A$ is adjacent to
another vertex in $A$.  If the graph is a lattice, the animal is termed a
lattice animal, and on an infinite lattice, the lattice animals are defined up
to an overall lattice translation.  A directed lattice animal $A_{dir}$ is an
animal on a lattice $\Lambda$ together with a special origin or root point $O$
such that any vertex of $A$ can be reached starting from $O$ by an oriented
path on $\Lambda$ having all of its vertices in $A$.  For example, a
directed lattice
animal on the square lattice could be defined by restricting the orientations
of each step of the oriented path to be ``eastward'' or ``northward'', so that,
starting at the origin, this animal would extend in the northeast quadrant and
would not contain any backtracking steps going in the south or west directions.
(The animals that we consider here are also called site animals to distinguish
them from bond animals; we shall leave this implicit.)  Animals and directed
animals are related, respectively, to percolation and directed percolation. 
Let us denote $N_{A,\Lambda,n}$ and $N_{DA,\Lambda,n}$ as the total number of
lattice animals and directed lattice animals with $n$ vertices, respectively,
on the lattice $\Lambda$.  For $n \to \infty$, the numbers of lattice animals
and directed lattice animals grow asymptotically like $n^{-\theta}a^n$, where 
$a$ depends on the lattice and, e.g., for 2D lattices, $\theta=1/2$
(e.g. \cite{dhar1}-\cite{bousq} and references therein).  Having given
this background, we now state the relations that we have obtained.

First, we find that the total number of distinct eigenvalues of the coloring
matrix, i.e. the total number, $N_{P,L_y,\lambda}$, of different
$\lambda_{P,G,j}$'s in the chromatic polynomial (\ref{pgsum}) for cyclic strips
with width $L_y$ of the square and triangular lattice (equal by (\ref{ncycmb})
to the same number for the corresponding M\"obius strips of the square and
triangular lattices) is twice the number of directed lattice animals with
$n=L_y$ vertices on the square (sq) lattice: 
\beq 
N_{P,L_y,\lambda}=2N_{DA,sq,L_y} \ .
\label{nptotdasq}
\eeq 
We established this by observing that the generating function for directed
lattice animals on the square lattice, which is known exactly
\cite{dhar1,dhar23}, is precisely $(1/2)$ times the generating function
(\ref{nptotgenfun}).

Second, we find that the analogous total number, $N_{Z,L_y,\lambda}$, of
different eigenvalues $\lambda_{Z,G,j}$'s appearing in the full Potts model
partition function (\ref{zgsum}) for these cyclic and M\"obius strips with
width $L_y$ of the square and triangular lattices is twice the number of
directed lattice animals with $n=L_y$ vertices on the triangular lattice:
\beq
N_{Z,L_y,\lambda}=2N_{DA,tri,L_y} \ . 
\label{nztotdatri}
\eeq 
We established this by recalling the known result \cite{dhar1,dhar23} 
\beq
N_{DA,tri,n}=\frac{1}{2} {2n \choose n} 
\label{ndatri}
\eeq
and using our calculation (\ref{nztot}). 
It should be emphasized that, as we have shown above, 
the numbers $N_{P,L_y,\lambda}$ and $N_{Z,L_y,\lambda}$ on the respective 
left-hand sides of (\ref{nptotdasq}) and (\ref{nztotdatri}) apply 
for cyclic and M\"obius strips of both the square
and triangular lattice, whereas the right-hand side of (\ref{nptotdasq}) is
specific to the square lattice and the right-hand side of (\ref{nztotdatri}) is
specific to the triangular lattice.

Parenthetically, we mention some other relations.  The enumeration of directed
lattice animals on the square and triangular lattices was shown in
\cite{dhar23} to be connected with the hard-square lattice gas model
\cite{baxtersq}.  As discussed in \cite{gutt2,bousq}, no
exact result has been obtained, analogous to those in \cite{dhar1,dhar23}, for
the number of directed lattice animals on the honeycomb lattice. 
Directed lattice animals are connected with directed percolation, and the
latter has been related to a kind of chiral Potts model \cite{essam}.  Compact
lattice animals are the subset of lattice animals with no unoccupied interior
sites; directed compact lattice animals have been related to the 
$q \to \infty$ limit of the Potts model and have been shown to have numbers
that grow less rapidly than the leading $a^n$ growth in the numbers of regular 
directed lattice animals \cite{wuetal}.

\section{Determination of $n_{P,Mb}(L_y,d,\pm)$ for M\"obius 
Strips of the Square Lattice}

M\"obius strips differ in their global topology from cyclic strips.  For the
M\"obius strips of the square lattice that have been studied, the coefficients
are polynomials in $q$, but the polynomials that occur for the strip of width
$L_y$ arise from the set $\pm c^{(d)}$, where $0 \le d \le L_y/2$ for even
$L_y$ and $0 \le d \le (L_y+1)/2$ for odd $L_y$, rather than the set $c^{(d)}$,
$0 \le d \le L_y$ as in the case of the cyclic strips.  In passing, we note
that for M\"obius strips of the triangular lattice, we have found, originally
for width $L_y=2$ in \cite{wcy} and later for width $L_y=3$ in \cite{t}, that
in the case of chromatic polynomials, some of the coefficients are not
polynomials, but rather, algebraic functions of $q$ \cite{wcy,t}; similarly,
some of the coefficients in the full Potts model partition function for the
M\"obius strip of the triangular lattice are not polynomials in $q$ \cite{ta}.
This is related to the fact, discussed before \cite{t,ta}, that the M\"obius
strip of the triangular lattice involves a seam, i.e., it is not
translationally homogeneous in the longitudinal direction, whereas the M\"obius
strip of the square lattice does not have a seam and is translationally
homogeneous in the longitudinal direction.  Hence, for M\"obius strips, we only
consider the square lattice here.

Since both signs of the $c^{(d)}$ occur, it is necessary to define a larger set
of numbers for M\"obius strips of the square lattice.  Let 
$n_{P,Mb}(L_y,d,\pm)$ denote the number of terms $\lambda_j$'s in 
the expression (\ref{pgsum}) for $P(sq,L_y \times L_x,FBC_y,TPBC_x,q)$ with 
coefficients $c_{P,L_y,Mb,j}=\pm c^{(d)}$, respectively, where $Mb$ denotes
M\"obius.  The notation $c_{P,L_y,Mb,j}$ reflects the fact that these
coefficients depend on $L_y$ and the M\"obius boundary conditions, but are 
independent of $L_x$. The sums and differences for each $d$ are defined as 
\beq
n_{P,Mb,tot}(L_y,d)= n_{P,Mb}(L_y,d,+)+n_{P,Mb}(L_y,d,-)
\label{nlydsum}
\eeq
and
\beq
\Delta n_{P,Mb}(L_y,d) = n_{P,Mb}(L_y,d,+)-n_{P,Mb}(L_y,d,-) \ .
\label{deltan}
\eeq
We first obtain a general theorem for the sum of the coefficients: 

\begin{flushleft}

Theorem 5.  

\beq
C_{P,L_y,Mb} = \sum_{j=1}^{N_{P,L_y,\lambda}} c_{P,L_y,Mb,j} = 
\sum_{d=0}^{d_{max}} \Delta n_{P,Mb}(L_y,d) c^{(d)} =
\cases{ 0 & for even $L_y$ \cr
P(T_{(\frac{L_y+1}{2})},q) & for odd $L_y$ \cr }
\label{csummb}
\eeq
where $P(T_n,q)$ was given in (\ref{pt}) and 
\beq
 d_{max}=  \cases{ \frac{L_y}{2} & for even $L_y$ \cr
\frac{(L_y+1)}{2} & for odd $L_y$ \cr } \ . 
\label{dmax}
\eeq

\vspace{6mm}

Proof.  \ Our method for proving this theorem is inspired by coloring matrix
methods \cite{matmeth}.  The M\"obius strip involves a reversed-orientation
periodic longitudinal boundary condition.  We can think of constructing such a
strip by cutting a cyclic strip, reversing the orientation of one of the ends,
and gluing these ends together again.  For a strip with $L_y$ even, let us
label the vertices on the two ends as $1,2,..,L_y$; then the M\"obius boundary
condition means identifying vertex 1 with vertex $L_y$, vertex 2 with vertex
$L_y-1$, and so forth.  As regards the coloring constraint, in the case
$L_y=2$, this effectively produces a subgraph consisting of a single vertex
with an edge forming a loop.  In the case $L_y=4$, this produces a vertex $1=4$
with two edges connecting to a vertex $2=3$, which in turn has a loop connected
to it, and so forth for higher even values of $L_y$.  Because of the loop that
appears in each even-$L_y$ case, the chromatic polynomial for coloring this
subgraph vanishes identically.  (The value of $d_{max}$ given in (\ref{dmax})
follows from this graphical construction.)  This proves the theorem for the
case of even $L_y$.  For odd $L_y$, consider first the case $L_y=3$; here the
M\"obius boundary condition identifies vertex 1 with vertex 3 and leaves vertex
2 invariant.  Hence, as regards the coloring, it effectively produces a
subgraph consisting of the vertex $1=3$ connected by two edges with the vertex
2.  The chromatic polynomial for the coloring of this subgraph is not sensitive
to the multiple edges and hence is $P(T_2,q)=q(q-1)$.  Again, this subgraph
construction yields the value of $d_{max}$ for odd $L_y$ in (\ref{dmax}).  In a
similar manner, for higher odd values of $L_y$, the M\"obius boundary condition
leads to a subgraph with $(L_y+1)/2$ vertices forming a chain, with each
interior pair connected to the next by two edges.  The chromatic polynomial for
the coloring of this subgraph is $P(T_{(L_y+1)/2},q)=q(q-1)^{(L_y-1)/2}$.  This
completes the proof of the theorem. $\Box$

\vspace{8mm}

Two corollaries of this theorem are as follows. 

\vspace{4mm}

Corollary 1 

\vspace{4mm}

\beq
n_{P,Mb}(L_y,d,+)=n_{P,Mb}(L_y,d,-) \quad {\rm for \ even} \quad L_y
\label{nmblyeven}
\eeq

\vspace{6mm} 

Proof. \ If $L_y$ is even, then in order for the terms of the highest power in
$q$ to cancel so as to yield a sum of 0 as in (\ref{csummb}), it is necessary
that $n_{P,Mb}(L_y,d_{max},+)=n_{P,Mb}(L_y,d_{max},-)$, where $d_{max}$ was
given above in (\ref{dmax}).  But then in turn, in order for the terms of
degree $d_{max}-1$ to sum to zero, it is necessary that
$n_{P,Mb}(L_y,d_{max}-1,+)=n_{P,Mb}(L_y,d_{max}-1,-)$, and so forth for all
powers. \ $\Box$

\vspace{6mm}

Corollary 2 

\beq
\Delta n_{P,Mb}(L_y,d)=n_{P}(\frac{L_y+1}{2},d) \quad {\rm for \ odd} \quad 
L_y \ge 3 
\label{deltanmblyodd}
\eeq

\vspace{6mm}

Proof.  \ This follows from eq. (\ref{csummb}) by the same kind of argument
that was used to obtain the $n_P(L_y,d)$.  Given that the sum 
$C_{P,L_y,Mb}$
satisfies (\ref{csummb}), this uniquely determines the $\Delta
n_{P,Mb}(L_y,d)$'s just as (\ref{ncrelcyc}) determined the $n_P(L_y,d)$. \
$\Box$

\vspace{6mm}

In order to determine the numbers of coefficients $c^{(d)}$, i.e.,
$n_{P,Mb}(L_y,d,\pm)$, for a M\"obius strip of width $L_y$, we start with the
chromatic polynomial for the cyclic $L_y$ strip and determine how the 
coefficients $c^{(d)}$ change when one changes the longitudinal boundary
condition from cyclic to M\"obius.  The next two theorems determines this (the
proofs are given after the second of these two theorems). 

\vspace{8mm}

Theorem 6 \  Consider an $L_y \times L_x$ strip of the square lattice with 
$(FBC_y,PBC_x)$.  As before, the coefficients $c_{P,G,j}$ are made up from
the set
$c^{(d)}$ with $0 \le d \le L_y$.  When one changes the longitudinal boundary
condition from cyclic to M\"obius, the following respective changes of 
coefficients of even degree $d=2k$ and of odd degree $2k+1$ occur:
\beq
c^{(0)} \to \pm c^{(0)}
\label{cd0tran}
\eeq
\beq
c^{(2k)} \to \pm c^{(k-1)} \ ,  1 \le k \le 
\Bigl [ \frac{L_y}{2} \Bigr ]
\label{cdeventran}
\eeq
\beq
c^{(2k+1)} \to \pm c^{(k+1)} \ ,  0 \le k \le
\Bigl [ \frac{L_y-1}{2} \Bigr ]
\label{cdoddtran}
\eeq 
where in eqs. (\ref{cdeventran}) and (\ref{cdoddtran}), we again use the 
notation $[\nu]$ to denote the integral part of $\nu$.  Thus, if there are 
$n_P(L_y,0)$ coefficients $c_{P,G,j}$ of the terms $(\lambda_j)^m$ of the form
$c^{(0)}$ in
eq. (\ref{pgsum}) for $P(sq(L_y \times L_x; FBC_y,PBC_x),q)$, then the
respective coefficients multiplying the terms $(\lambda_j)^m$ in the chromatic
polynomial for the corresponding M\"obius strip, $P(sq(L_y \times L_x;
FBC_y,TPBC_x),q)$, are either of the form $+c^{(0)}$ or $-c^{(0)}$, and so
forth for the $c^{(d)}$ with $d > 0$, as specified by eqs. (\ref{cdeventran})
and (\ref{cdoddtran}).  

\vspace{8mm}

Theorem 7. With the same premise as in Theorem 6, we have 

\beq
n_{P,Mb}(L_y,0,\pm) = \frac{1}{2}n_{P,Mb,tot}(L_y,0)=
\frac{1}{2}\biggl [ n_P(L_y,0)+n_P(L_y,2) \biggr ] \quad {\rm for } 
\quad L_y \ \ {\rm even}
\label{ntranlyevend0}
\eeq

\beqs
n_{P,Mb}(L_y,d,\pm) & = & \frac{1}{2}n_{P,Mb,tot}(L_y,d) 
= \frac{1}{2}\biggl [ n_P(L_y,2d-1)+n_P(L_y,2d+2) \biggr ] \cr\cr
& & \quad {\rm for } \quad L_y \ \ {\rm even} 
\quad {\rm and} \quad 1 \le d \le \frac{L_y}{2}
\label{ntranlyeven}
\eeqs

\beq
n_{P,Mb}(L_y,d,\pm) = 0 \quad {\rm for } \quad L_y \ \ {\rm even} 
\quad {\rm and} \quad d \ge \frac{L_y+2}{2}
\label{nzerolyeven}
\eeq

\beq
n_{P,Mb}(L_y,0,\pm) = \frac{1}{2}\biggl [ n_P(L_y,0)+n_P(L_y,2) \pm
n_P(\frac{L_y+1}{2},0) \biggr ] \quad {\rm for } \quad L_y \ \ {\rm odd} 
\label{ntranlyoddd0}
\eeq

\beqs
n_{P,Mb}(L_y,d,\pm) & = & \frac{1}{2}\biggl [ n_P(L_y,2d-1)+n_P(L_y,2d+2) \pm
n_P(\frac{L_y+1}{2},d) \biggr ] \cr\cr
& & \quad {\rm for } \quad L_y \ \ {\rm odd} 
\quad {\rm and} \quad 1 \le d \le \frac{L_y+1}{2}
\label{ntranlyodd}
\eeqs

\beq
n_{P,Mb}(L_y,d,\pm) = 0 \quad {\rm for } \quad L_y \ \ {\rm odd} 
\quad {\rm and} \quad d \ge \frac{L_y+3}{2}
\label{nzerolyodd}
\eeq

\vspace{8mm}

Proofs.  \ We prove these theorems using an inductive argument.  From the
chromatic polynomials for the cyclic and M\"obius strips of the square lattice
with width $L_y=2$ \cite{bds}, one knows the transformation rule for the
coefficients $c^{(d)}$ with $0 \le d \le 2$ as one goes from cyclic to M\"obius
longitudinal boundary conditions, namely, $c^{(0)} \to \pm c^{(0)}$, $c^{(1)}
\to \pm c^{(1)}$, and $c^{(2)} \to \pm c^{(0)}$, where the sign information
will not be needed.  For $L_y=3$, we know the numbers $n_P(L_y,d)$ for the
cyclic strip from the previous theorem.  We also know the differences of the
numbers of coefficients of each degree for the $L_y=3$ M\"obius strip, $\Delta
n_{P,Mb}(3,d)$; these are determined by the relation (\ref{csummb}).  In
particular, we have $\Delta n_{P,Mb}(3,0)=n_P(2,0)=1$, $\Delta
n_{P,Mb}(3,1)=n_P(2,1)=2$, and $\Delta n_{P,Mb}(3,2)=n_P(2,2)=1$. From the
transformation rules obtained so far, it follows that the sources for
coefficients of degree 0 for the $L_y=3$ M\"obius strip are the coefficients of
degree 0 and 2 for the corresponding cyclic strip.  Adding these, we have
$n_{P,Mb,tot}(3,0)=n_P(3,0)+n_P(3,2)=5$. This relation for the sum of
$n_{P,Mb}(3,0,+)$ and $n_{P,Mb}(3,0,-)$, together with the relation $\Delta
n_{P,Mb}(3,0)=1$ for their difference, enables us to solve for each of these
numbers, and we get $n_{P,Mb}(3,0,+)=3$, $n_{P,Mb}(3,0,-)=2$.  From the
transformation rule $c^{(1)} \to \pm c^{(1)}$ found previously, we infer that
for the M\"obius strip, $n_{P,Mb,tot}(3,1)=n_P(3,1)=4$.  Combining this with
the relation $\Delta n_{P,Mb}(3,1)=2$, we solve to get $n_{P,Mb}(3,1,+)=3$ and
$n_{P,Mb}(3,1,-)=1$.  We next consider $n_{P,Mb}(3,2,\pm)$.  We know that
$\Delta n_{P,Mb}(3,2)=1$, and, given the transformation rules obtained so far,
there is only one source for terms with degree coefficients of degree 2 in the
chromatic polynomial for the $L_y=2$ M\"obius strip, namely the single term
with a $c^{(3)}$ coefficient.  Hence, $n_{P,Mb,tot}(3,2)=1$, and so, knowing
the sum and difference, we get $n_{P,Mb}(3,2,+)=1$ and $n_{P,Mb}(3,2,-)=0$.
Since there are no sources for any higher-degree coefficients, we have
$n_{P,Mb}(3,d,\pm)=0$ for $d \ge 3$. Thus, we have both determined all of the
numbers $n_{P,Mb}(3,d,\pm)$ and the transformation rule for the coefficients of
next higher degree, namely $c^{(3)} \to \pm c^{(2)}$.

Proceeding to $L_y=4$, we start with the knowledge of the $n_P(4,d)$ for the
cyclic strip and of the differences $\Delta n_{P,Mb}(4,d)=0$. For
coefficients of
degree 0 in the chromatic polynomial for the M\"obius strip, the transformation
rules indicate two sources, namely the coefficients with degree 0 or 2 in the
corresponding cyclic strip.  We shall show, {it a posteriori}, that these are
the only sources, so we infer that $n_{P,Mb,tot}(4,0)=n_P(4,0)+n_P(4,2)=12$.
Combining this with the relation $\Delta n_{P,Mb}(4,0)=0$, we obtain
$n_{P,Mb}(4,0,\pm)=(1/2)n_{P,Mb,tot}(4,0)=6$.  Next, for the coefficients of
degree 1, we start with those from the cyclic strip, $n_P(4,1)=9$.  But we know
that there must be another source because this is an odd number and, by itself,
would give the unacceptable non-integral result for
$n_{P,Mb}(4,0,\pm)=(1/2)n_{P,Mb,tot}(4,0)$.  The only possible source is the
coefficients of degree 4 in the cyclic strip, so we infer the next
transformation rule: $c^{(4)} \to \pm c^{(1)}$.  This yields
$n_{P,Mb,tot}(4,1)=n_P(4,1)+n_P(4,4)=10$ and hence
$n_{P,Mb}(4,0,\pm)=5$. Finally, for the coefficients of degree 2 in the
chromatic polynomial for the $L_y=4$ M\"obius strip, the source is the
coefficients of degree 3 in the cyclic $L_y=4$ strip, so
$n_{P,Mb,tot}(4,2)=n_P(4,3)=4$, whence $n_{P,Mb}(4,2,\pm)=2$.  Since there are
no sources for any higher-degree coefficients, we have $n_{P,Mb}(4,d,\pm)=0$
for $d \ge 3$.  Thus again we have determined all of the numbers
$n_{P,Mb}(4,d,\pm)$ and also the next higher transformation rule for
$c^{(4)}$.

It should now be clear how one proceeds iteratively: at each higher width
$L_y$, one uses the previously established transformation rules, and obtains
the transformation rule for the next higher degree coefficient, to determine
all of the numbers $n_{P,Mb}(L_y,d,\pm)$.  This completes the proof of Theorems
5 and 6.  $\Box$. 

\end{flushleft}

Two corollaries of these theorems are
\beq
n_{P,Mb}(L_y,\frac{L_y}{2},\pm)=\frac{L_y}{2} \quad {\rm for \ even} 
\quad L_y
\label{ncdmbmaxlyeven}
\eeq
\beq
n_{P,Mb}(L_y,\frac{L_y+1}{2},+)=1 \ , \quad n_{P,Mb}(L_y,\frac{L_y+1}{2},-)=0 
\quad {\rm for \ odd} \quad L_y
\label{ncdmbmaxlyodd}
\eeq
Values of the numbers $n_{P,Mb}(L_y,d,\pm)$ for the first several widths, 
$2 \le L_y \le 10$, are given in Table \ref{npctablemb}. 

\begin{table}
\caption{\footnotesize{Table of numbers $n_{P,Mb}(L_y,d,\pm)$ for M\"obius
strips of width $L_y$.  For each $L_y$ value, the entries in the first and
second lines are $n_{P,Mb}(L_y,d,+)$ and $n_{P,Mb}(L_y,d,-)$,
respectively. Blank entries are zero. The last entry for each value of $L_y$ is
the total $N_{P,L_y,Mb,\lambda}$.}}
\begin{center}
\begin{tabular}{|c|c|c|c|c|c|c|c|}
\hline\hline
$L_y \ \downarrow$ \ \ $(d,+) \ \rightarrow$
   & $0,+$ & $1,+$ & $2,+$ & $3,+$ & $4,+$ & $5,+$ & \\
\quad \quad $(d,-) \ \rightarrow$ 
   & $0,-$ & $1,-$ & $2,-$ & $3,-$ & $4,-$ & $5,-$ 
& $N_{P,L_y,Mb,\lambda}$ \\ \hline\hline
2  & 1    & 1    &      &     &    &    &       \\ 
   & 1    & 1    &      &     &    &    & 4    \\ \hline
3  & 3    & 3    & 1    &     &    &    &      \\ 
   & 2    & 1    &      &     &    &    & 10   \\ \hline
4  & 6    & 5    & 2    &     &    &    &      \\
   & 6    & 5    & 2    &     &    &    & 26   \\ \hline
5  & 16   & 15   & 8    & 1   &    &    &      \\ 
   & 14   & 11   & 5    &     &    &    & 70   \\ \hline
6  & 38   & 35   & 20   & 3   &    &    &      \\
   & 38   & 35   & 20   & 3   &    &    & 192  \\ \hline
7  & 100  & 100  & 64   & 15  & 1  &    &      \\ 
   & 96   & 91   & 56   & 11  &    &    & 534  \\ \hline
8  & 256  & 263  & 178  & 49  & 4  &    &      \\ 
   & 256  & 263  & 178  & 49  & 4  &    & 1500 \\ \hline
9  & 681  & 739  & 535  & 178 & 24 & 1  &      \\ 
   & 672  & 718  & 514  & 165 & 19 &    & 4246 \\ \hline
10 & 1805 & 2027 & 1539 & 574 & 96 & 5  &      \\
   & 1805 & 2027 & 1539 & 574 & 96 & 5  & 12092 \\ 
\hline\hline
\end{tabular}
\end{center}
\label{npctablemb}
\end{table}

Further corollaries involve ($q$-independent) relations between the 
$n_P(L_y,d)$.  As before for the cyclic strips, we derive these by evaluating 
(\ref{csummb}) for $q=0,1$, and 2.  For example, 
setting $q=0$ and using (\ref{cdq0}), we have (for $L_y \ge 2$ where M\"obius
strips are defined) 
\beq
\sum_{d=0}^{d_{max}} (-1)^d \Delta n_{P,Mb}(L_y,d) = 0 
\quad {\rm where} \ \ d_{max}=  \cases{ \frac{L_y}{2} & for even $L_y$ \cr
\frac{(L_y+1)}{2} & for odd $L_y$ \cr }
\label{cdmbsumq0}
\eeq
It is straightforward to derive similar relations among the $\Delta
n_{P,Mb}(L_y,d)$ by evaluating (\ref{csummb}) for $q=1$ and $q=2$. 

\vspace{6mm}

Another result pertains to the detailed structure of the chromatic polynomials
for the cyclic, as compared with M\"obius strips of a given width $L_y$, as the
length $L_x$ gets large.  As discussed in \cite{pm}, because the cyclic and
M\"obius strips of a given width have the same number of vertices and edges,
the coefficients of the leading powers of $q$ are the same.  We recall this
result.  For $m$ greater than a minimal value\footnote{
\footnotesize{For example, for $m \le 4$  ($m \le 3$) the $L_y=2$ cyclic
(M\"obius) square strips exhibit special behavior regarding $g$ and $k_g$;
for $m$ larger than these respective values, they both have $g=4$ and
$k_g=m$.}}, the cyclic and M\"obius strips of a given $(G_s)_m$ have the same
number of vertices $n$, edges (bonds) $e$, girth $g$ (length of minimum
closed circuit on $G_s$) and number $k_g$ of circuits of length $g$.  One has
$n=t_s m$ where $t_s$ depends on $G_s$. Writing
\beq
P((G_s),q) = \sum_{j=0}^{n-1}(-1)^j h_{n-j}q^{n-j}
\label{p}
\eeq
and using the results that \cite{rtrev,m} 
$h_{n-j}={e \choose j}$ for $0 \le j < g-1$ (whence $h_n=1$ and $h_{n-1}=e$)
and $h_{n-(g-1)}={e \choose g-1}-k_g$, it follows that for $m$ greater than the
above-mentioned minimal value, these $h_j$'s are the same for the cyclic and
M\"obius strips of each type $G_s$.  For a given $G_s$, as $m$ increases, the
$h_{n-j}$'s for the cyclic and M\"obius strips become equal for larger $j$.
Our relations (\ref{cd0tran})-(\ref{cdoddtran}) make clear which coefficients
$c_{G_s(L_y),j}$ in (\ref{pgsum}), and hence which coefficients 
$h_{G_s(L_y),j}$ in (\ref{p}), remain equal for the cyclic versus M\"obius
strips.

\section{Determination of the numbers $n_{Z,Mb}(L_y,d,\pm)$ for M\"obius 
Strips of the Square Lattice}

In this section we generalize the calculations of the previous section to the
full Potts model partition function on M\"obius strips of the square lattice. 
(The restriction to the square lattice will be implicit in the notation.) 
Let $n_{Z,Mb}(L_y,d,\pm)$ denote the number of terms $\lambda_{Z,L_y,j}$ in
the expression (\ref{zgsum}) for $Z(L_y \times L_x,FBC_y,TPBC_x,q,v)$ with
coefficients $c_{Z,L_y,Mb,j}=\pm c^{(d)}$, respectively, where $Mb$ denotes
M\"obius.  The notation $c_{Z,L_y,Mb,j}$ reflects the fact that these
coefficients are independent of $L_x$. The sums and differences for each $d$ 
are defined as
\beq
n_{Z,Mb,tot}(L_y,d)= n_{Z,Mb}(L_y,d,+)+n_{Z,Mb}(L_y,d,-)
\label{nlydsumz}
\eeq
and
\beq
\Delta n_{Z,Mb}(L_y,d) = n_{Z,Mb}(L_y,d,+)-n_{Z,Mb}(L_y,d,-) \ . 
\label{deltanz}
\eeq

We first have the two theorems 

\begin{flushleft}

Theorem 8.

\beq
C_{Z,L_y,Mb} \equiv \sum_{j=1}^{N_{Z,L_y,\lambda}} c_{Z,L_y,Mb,j} =
\sum_{d=0}^{d_max} \Delta n_{Z,Mb}(L_y,d) c^{(d)} =\cases{ q^{L_y/2} & for
even $L_y$ \cr
q^{(L_y+1)/2} & for odd $L_y$ \cr }
\label{ctsummb}
\eeq

\vspace{6mm}

Theorem 9. \ For the M\"obius strips of the square lattice, 
\beq
n_{Z,Mb}(L_y,0,\pm) = 
\frac{1}{2}\biggl [ n_Z(L_y,0)+n_Z(L_y,2)\pm n_Z(\frac{L_y}{2},0) \biggr ] 
\quad {\rm for } \quad L_y \ \ {\rm even}
\label{ntranlyevend0mb}
\eeq

\beqs
n_{Z,Mb}(L_y,d,\pm) & = & 
\frac{1}{2}\biggl [ n_Z(L_y,2d-1)+n_Z(L_y,2d+2)\pm n_Z(\frac{L_y}{2},d) 
\biggr ] \cr\cr
& & \quad {\rm for } \quad L_y \ \ {\rm even}
\quad {\rm and} \quad 1 \le d \le \frac{L_y}{2}
\label{ntranlyevenmb}
\eeqs

\beq
n_{Z,Mb}(L_y,0,\pm) =
\frac{1}{2}\biggl [ n_Z(L_y,0)+n_Z(L_y,2)\pm n_Z(\frac{L_y+1}{2},0) \biggr ]
\quad {\rm for } \quad L_y \ \ {\rm odd}
\label{ntranlyodd0mb}
\eeq

\beqs
n_{Z,Mb}(L_y,d,\pm) & = & 
\frac{1}{2}\biggl [ n_Z(L_y,2d-1)+n_Z(L_y,2d+2)\pm n_Z(\frac{L_y+1}{2},d)
\biggr ] \cr\cr
& & \quad {\rm for } \quad L_y \ \ {\rm odd}
\quad {\rm and} \quad 1 \le d \le \frac{L_y+1}{2} \ . 
\label{ntranlyoddmb}
\eeqs

\beq
n_{Z,Mb}(L_y,d,\pm)=0 \quad {\rm for} \quad d > \Bigl [ \frac{L_y+1}{2} \Bigr ]
\label{ntmblimit}
\eeq

\vspace{6mm}

The proofs of these theorems are analogous to those for the analogous Theorems
5 and 7 for chromatic polynomials for M\"obius strips of the square lattice. 

\end{flushleft}

Two corollaries of these theorems are
\beq
n_{Z,Mb}(L_y,\frac{L_y}{2},+)=L_y \ , \quad 
n_{Z,Mb}(L_y,\frac{L_y}{2},-)=L_y-1 \quad {\rm for \ even} \quad L_y
\label{nctmbmaxlyeven}
\eeq

\beq
n_{Z,Mb}(L_y,\frac{L_y+1}{2},+)=1 \ , \quad n_{Z,Mb}(L_y,\frac{L_y+1}{2},-)=0
\quad {\rm for \ odd} \quad L_y
\label{nctmbmaxlyodd}
\eeq
Values of the numbers $n_{Z,Mb}(L_y,d,\pm)$ for the first several widths,
$2 \le L_y \le 10$, are given in Table \ref{ntctablemb}.

\begin{table}
\caption{\footnotesize{Table of numbers $n_{Z,Mb}(L_y,d,\pm)$ for M\"obius
strips of width $L_y$.  For each $L_y$ value, the entries in the first and
second lines are $n_{Z,Mb}(L_y,d,+)$ and $n_{Z,Mb}(L_y,d,-)$,
respectively. Blank entries are zero. The last entry for each value of $L_y$ is
the total $N_{Z,L_y,Mb,\lambda}$.}}
\begin{center}
\begin{tabular}{|c|c|c|c|c|c|c|c|}
\hline\hline
$L_y \ \downarrow$ \ \ $(d,+) \ \rightarrow$
   & $0,+$ & $1,+$ & $2,+$ & $3,+$ & $4,+$ & $5,+$ & \\
\quad \quad $(d,-) \ \rightarrow$
   & $0,-$ & $1,-$ & $2,-$ & $3,-$ & $4,-$ & $5,-$
& $N_{Z,L_y,Mb,\lambda}$ \\ \hline\hline
2  & 2     & 2     &       &      &     &    &      \\
   & 1     & 1     &       &      &     &    & 6    \\ \hline
3  & 6     & 6     & 1     &      &     &    &      \\ 
   & 4     & 3     &       &      &     &    & 20   \\ \hline
4  & 18    & 16    & 4     &      &     &    &      \\
   & 16    & 13    & 3     &      &     &    & 70   \\ \hline
5  & 61    & 54    & 20    & 1    &     &    &       \\
   & 56    & 45    & 15    &      &     &    & 252   \\ \hline
6  & 206   & 180   & 80    & 6    &     &    &       \\
   & 201   & 171   & 75    & 5    &     &    & 924   \\ \hline
7  & 722   & 651   & 335   & 42   & 1   &    &       \\
   & 708   & 623   & 315   & 35   &     &    & 3432  \\ \hline
8  & 2542  & 2360  & 1336  & 224  & 8   &    &       \\
   & 2528  & 2332  & 1316  & 217  & 7   &    & 12870 \\ \hline
9  & 9082  & 8766  & 5367  & 1148 & 72  & 1  &        \\
   & 9040  & 8676  & 5292  & 1113 & 63  &    & 48620  \\ \hline
10 & 32644 & 32668 & 21270 & 5432 & 480 & 10 &        \\
   & 32602 & 32578 & 21195 & 5397 & 471 & 9  & 184756 \\ \hline\hline
\end{tabular}
\end{center}
\label{ntctablemb}
\end{table}

Two relations among the $n_{Z,Mb}(L_y,d)$ can be derived as before, by
evaluating the sum (\ref{ctsummb}) for $q=0$ and $q=1$.  For example, from the
evaluation for $q=0$ we have (for $L_y\ge 2$)
\beq
\sum_{d=0}^{d_{max}} (-1)^d (\Delta n_{Z,Mb}(L_y,d))=0 
\quad {\rm where} \ \ d_{max}=  \cases{ \frac{L_y}{2} & for even $L_y$ \cr
\frac{(L_y+1)}{2} & for odd $L_y$ \cr }
\label{ctsummbq0}
\eeq
It is straightforward to obtain a corresponding relation via the evaluation of
(\ref{ctsummb}) for $q=1$.

\section{Some Examples}

Since some of the notation above is complicated, it is worthwhile to illustrate
our general calculations with specific exact solutions.  These will also be
useful in a discussion of the special cases of the infinite-temperature limit
of the Potts model and the zero-temperature limit of the Potts ferromagnet
below. 

\subsection{Chromatic Polynomials}

For the $L_y=3$ cyclic
strip of the square lattice, $N_{P,L_y=3,\lambda}=10$ and, with the
abbreviation $sq3FP$ for $sq,L_y=3,FBC_y,PBC_x$, the chromatic polynomial is
\cite{wcy} 
\beq 
P(sq,3 \times L_x,FBC_y,PBC_x,q)=\sum_{j=1}^{10}
c_{sq3FP,j}(\lambda_{sq3FP,j})^{L_x}
\label{psq3fp}
\eeq
where
\beq
\lambda_{sq3FP,1}=-1
\label{lamsq3cyc1}
\eeq
\beq
\lambda_{sq3FP,2}=q-1
\label{lamsq3cyc2}
\eeq
\beq
\lambda_{sq3FP,3}=q-2
\label{lamsq3cyc3}
\eeq
\beq
\lambda_{sq3FP,4}=q-4
\label{lamsq3cyc4}
\eeq
\beq
\lambda_{sq3FP,5}=-(q-2)^2
\label{lamsq3cyc5}
\eeq
\beq
\lambda_{sqFP,(6,7)} = \frac{1}{2}\Biggl [ (q-2)(q^2-3q+5) \pm
\Bigl \{ (q^2-5q+7)(q^4-5q^3+11q^2-12q+8) \Bigr \}^{1/2} \Biggr ]
\label{lamsq3cyc67}
\eeq
and $\lambda_{sqFP,j}$, $j=8,9,10$, are the roots of the cubic equation
\beqs
& & \xi^3+(2q^2-9q+12)\xi^2+(q^4-10q^3+36q^2-56q+31)\xi \cr\cr
& & -(q-1)(q^4-9q^3+29q^2-40q+22)=0 \ . 
\label{sqcubic}
\eeqs
The coefficients are
\beq
c_{sq3FP,1}=c^{(3)}
\label{csq3cyc1}
\eeq
\beq
c_{sq3FP,j}=c^{(2)} \quad {\rm for} \ \ j=2,3,4
\label{csq3cyc234}
\eeq
\beq
c_{sq3FP,j}=c^{(1)} \quad {\rm for} \ \ j=5,8,9,10
\label{csq3cyc810}
\eeq
\beq
c_{sq3FP,j}=c^{(0)}=1 \quad {\rm for} \ \ j=6,7 \ . 
\label{csq3cyc67}
\eeq
For the numbers $n_P(L_y,d)$, one sees that $n_P(3,0)=2$, $n_P(3,1)=4$, 
$n_P(3,2)=3$, $n_P(3,3)=1$, and the total number $N_{P,L_y,\lambda}=10$,
in agreement with Table \ref{npctablecyc}.  The exact solution for the 
chromatic polynomial for the $L_y=3$ cyclic strip of the triangular lattice,
given in \cite{t}, has the same $N_{P,L_y,\lambda}$ and $n_P(L_y,d)$'s, 
although eight of the ten $\lambda_{P,G,j}$'s are different. 

\vspace{6mm}

As an illustration of our results for chromatic polynomials for M\"obius 
strips of the square lattice, we again take the $L_y=3$ case.  The exact
solution for the chromatic polynomial (with the 
abbreviation $sq3Mb$ for $sq,L_y=3,FBC_y,TPBC_x$) is \cite{pm}
\beq
P(sq,3 \times L_x,FBC_y,TPBC_x,q)=\sum_{j=1}^{10} 
c_{P,sq3Mb,j}(\lambda_{P,sq3Mb,j})^{L_x}
\label{psq3ftp}
\eeq
where $\lambda_{P,sq3Mb,j}=\lambda_{P,sq3FP,j} \ \ \forall \ j$, in accordance
with the general result (\ref{lameqcycmb}) and 
\beq
c_{P,sq3Mb,1}=c^{(2)}
\label{csq3mb}
\eeq
\beq
c_{P,sq3Mb,j}=c^{(1)} \quad {\rm for} \ \ j=8,9,10
\label{csq3mb810}
\eeq
\beq
c_{P,sq3Mb,5}=-c^{(1)}
\label{csq3mb5}
\eeq
\beq
c_{P,sq3Mb,j}=c^{(0)} \quad {\rm for} \ \ j=3,6,7
\label{csq3mb367}
\eeq
\beq
c_{P,sq3Mb,j}=-c^{(0)} \quad {\rm for} \ \ j=2,4 \ . 
\label{csq3mb24}
\eeq
Thus, for the numbers $n_P(L_y,d,\pm)$, we have $n_P(3,0,+)=3$, 
$n_P(3,0,-)=2$, $n_P(3,1,+)=3$, $n_P(3,1,-)=1$, $n_P(3,2,+)=1$, and 
$n_P(3,2,-)=0$, with $n_P(3,d,\pm)=0$ for $d \ge 3$, in agreement with Table
\ref{npctablemb}. 

The fact that the coefficients in the chromatic polynomial for the M\"obius
strip of the triangular lattice are not, in general, polynomials in $q$, was 
shown via the exact solution for the lowest width, $L_y=2$ in \cite{wcy} (with
$L_x=m$): 
\beqs
& & P(tri,2 \times L_x,FBC_y,TPBC_x,q) = -1 + [(q-2)^2]^m
- \frac{(q-1)(q-3)}{\sqrt{9-4q}}\biggl [
(\lambda_{t2Mb,3})^m - (\lambda_{t2Mb,4})^m \biggr ] \cr\cr
& = & -1 + (q-2)^{2m} -
2^{1-m}(q-1)(q-3)\sum_{s=0}^{[(m-1)/2]}{{m}\choose{2s+1}}(5-2q)^{m-2s-1}
(9-4q)^s
\label{pttly2pol}
\eeqs
where
\beq
\lambda_{t2Mb,(3,4)}= \frac{1}{2}\biggl [ 5-2q \pm \sqrt{9-4q} \ \biggr ]
\ .
\label{lambda34tly2cyc}
\eeq
As we have discussed in \cite{t}, this can be attributed to the feature that
the M\"obius strip of the triangular lattice has a seam, and hence is not
translationally homogeneous in the longitudinal direction, in contrast to the
M\"obius strip of the square lattice, which is translationally homogeneous in
the longitudinal direction. 

\subsection{Potts Model Partition Function} 

As an illustration for the full Potts model partition function, we recall the 
calculation of this function for the $L_y=2$ cyclic strip of the square
lattice.  Here, for $N_{Z,L_y,\lambda}$, we have \cite{bcc,a} 
$N_{Z,2,\lambda}=6$, and (with the shorthand $sq2FP$ for 
$sq,L_y=2,FBC_y,PBC_x$) 
\beq
Z(sq,2 \times L_x,FBC_y,PBC_x,q,v)=\sum_{j=1}^6
c_{Z,sq2FP,j}(\lambda_{Z,sq2FP,j})^{L_x} 
\label{zsq2cyc}
\eeq
where 
\beq
\lambda_{Z,sq2FP,1} = v^2
\label{lam1}
\eeq
\beq
\lambda_{Z,sq2FP,2} = v(v+q)
\label{lam2}
\eeq
\beq
\lambda_{Z,sq2FP,(3,4)} = \frac{v}{2}\Bigl [ q+v(v+4) \pm
(v^4+4v^3+12v^2-2qv^2+4qv+q^2)^{1/2} \Bigr ]
\label{lam34}
\eeq
\beq
\lambda_{Z,sq2FP,(5,6)} = \frac{1}{2}(T_{S12} \pm \sqrt{R_{S12}} \ )
\label{lams}
\eeq
with
\beq
T_{S12}=v^3+4v^2+3qv+q^2
\label{t56}
\eeq
\beq
R_{S12}=v^6+4v^5-2qv^4-2q^2v^3+12v^4+16qv^3+13q^2v^2+6q^3v+q^4 \ .
\label{rs12}
\eeq
and
\beq
c_{Z,sq2FP,1}=c^{(2)}
\label{clad1}
\eeq
\beq
c_{Z,sq2FP,j}=c^{(1)} \quad {\rm for} \ \ j=2,3,4
\label{clad234}
\eeq
\beq
c_{Z,sq2FP,j}=c^{(0)} \quad {\rm for} \ \ j=5,6 \ . 
\label{clad56}
\eeq
Thus, for the individual numbers, $n_Z(2,0)=2$, $n_Z(2,1)=3$, $n_Z(2,2)=1$, and
$n_Z(2,d)=0$ for $d \ge 3$, in agreement with Table \ref{ntctablecyc}. 

As an example of our general results for the structure of the
Potts model partition function on M\"obius strips of the square lattice, we
display the exact solution for width $L_y=2$ \cite{bcc,a} 
(with the abbreviation $sq2Mb$ for $sq,L_y=2,FBC_y,TPBC_x$):
\beq
Z(sq,2 \times L_x,FBC_y,TPBC_x,q,v)=\sum_{j=1}^6
c_{Z,sq2Mb,j}(\lambda_{Z,sq2Mb,j})^{L_x}
\label{zsq2mb}
\eeq
where $\lambda_{Z,sq2Mb,j}=\lambda_{Z,sq2FP,j} \ \ \forall \ j$, in accord with
the result (\ref{lamzeqcycmb}) and 
\beq
c_{Z,sq2Mb,j}=c^{(1)} \quad {\rm for} \ \ j=3,4
\label{cladmb34}
\eeq
\beq
c_{Z,sq2Mb,2}=-c^{(1)}
\label{cladmb2}
\eeq
\beq
c_{Z,sq2Mb,j}=c^{(0)} \quad {\rm for} \ \ j=5,6
\label{cladmb56}
\eeq
\beq
c_{Z,sq2Mb,1}=-c^{(0)}
\label{cladmb1}
\eeq
Thus, $n_{Z,Mb}(2,0,+)=2$, $n_{Z,Mb}(2,0,-)=1$, $n_{Z,Mb}(2,1,+)=2$,
and $n_{Z,Mb}(2,1,+)=1$, in agreement with Table \ref{ntctablemb}. 

\subsection{Other Limits}

In the text above we have given results for structural properties of the
general Potts model partition function $Z(G,q,v)$ and for the special case of
the zero-temperature Potts antiferromagnet, $Z(G,q,v=-1)=P(G,q)$.  Another
special case is infinite-temperature, i.e., $v=0$. In this case, for any graph
$G$, the Potts model partition function reduces to $Z(G,q,0)=q^{n(G)}$, as
indicated above in (\ref{zvzero}).  For the strip graphs under consideration
here, the equality in (\ref{zvzero}) corresponds to the property that at $v=0$
all but one of the $\lambda_{Z,G,j}$'s in eq. (\ref{zgsum}) vanish, and the
nonvanishing $\lambda_{Z,G,j}$ necessarily has the value $q^{L_y}$ and the
coefficient $c^{(0)}=1$, so that the right-hand side of (\ref{zgsum}) becomes
$q^{L_yL_x}=q^n$.  One can see this explicitly in the exact solution given
above for the Potts model partition function for the $L_y=2$ strip of the
square lattice.

Secondly, one may consider the zero-temperature limit of the Potts ferromagnet,
$v \to \infty$.  This is quite different from the $T=0$ limit of the Potts
antiferromagnet; among other differences, the ground state entropy per site is
zero for the ferromagnet rather than being nonzero as it is for the 
antiferromagnet with sufficiently large $q$.  
In this limit $T \to 0$, the partition function of the Potts
ferromagnet grows like $a^{e(G)}$ where $a$ was defined in
(\ref{kdef}) and $e(G)$ was also defined in the introduction as the number of
edges (bonds) in $G$. Hence, it is convenient to use the reduced partition 
function $Z_r$, defined by 
\beq 
Z_r(G,q,v)=a^{-e(G)}Z(G,q,v)=u^{e(G)}Z(G,q,v)
\label{zr}
\eeq
which has the finite limit $Z_r \to 1$ as $T \to 0$.  For the cyclic or
M\"obius strip graphs of the lattice $\Lambda$ of interest here we thus write
\beqs
Z_r(\Lambda,FBC_y,(T)PBC_x,q,v) & = & 
u^{e(G)}\sum_{j=1}^{N_{Z,G,\lambda}} c_{Z,G,j}
(\lambda_{Z,G,j})^{L_x} \cr\cr & \equiv & \sum_{j=1}^{N_{Z,G,\lambda}} 
c_{Z,G,j} (\lambda_{Z,G,j,u})^{L_x} 
\label{zu}
\eeqs
where as before we use the shorthand notation $G=\Lambda,FBC_y,(T)PBC_x$, and 
\beq
\lambda_{Z,G,j,u}=u^{e(G)/L_x}\lambda_{Z,G,j} \ . 
\label{lamu}
\eeq
For example, for cyclic or M\"obius strips of the square lattice, 
$e=(2L_y-1)L_x$, so the prefactor in (\ref{lamu}) is $u^{2L_y-1}$, while for
the corresponding strips of the triangular lattice, 
$e=(3L_y-2)L_x$, so the prefactor is $u^{3L_y-2}$.  For the $L_y=1$ 
circuit graph, 
\beq
\lambda_{Z,C,1,u}=1-u \ , \quad \lambda_{Z,C,2,u}=1+(q-1)u \ .
\label{lamredcirc}
\eeq
so that both of these two $\lambda_{Z,G,j,u}$'s remain important in the $T=0$
limit.  In contrast, for the $L_y=2$ cyclic or M\"obius strips of the square
and triangular lattices, our exact solutions in \cite{a,ta} show that, of the
six $\lambda_{Z,G,j,u}$'s, two are finite and nonzero for $u \to 0$ while the
other four vanish.  For example, for the cyclic/M\"obius strip of the square
lattice, in the vicinity of the zero-temperature point $u=0$, one has \cite{a}
\beq
\lambda_{Z,sq2FP,1,u}=u-2u^2+u^3
\label{lam1rtaylor}
\eeq
\beq
\lambda_{Z,sq2FP,2,u}=u+(q-2)u^2+(1-q)u^3
\label{lam2rtaylor}
\eeq
and the Taylor series expansions
\beq
\lambda_{Z,sq2FP,3,u}=1-u^2+2(q-2)u^3+O(u^4)
\label{lam3rtaylor}
\eeq
\beq
\lambda_{Z,sq2FP,4,u}=u+(q-4)u^2+(7-3q)u^3+O(u^4)
\label{lam4rtaylor}
\eeq
\beq
\lambda_{Z,sq2FP,5,u}=1+(q-1)u^2\Bigl [ 1 + 4u + O(u^2) \Bigr ]
\label{lam5rtaylor}
\eeq
\beq
\lambda_{Z,sq2FP,6,u}=u+2(q-2)u^2+(q^2-7q+7)u^3 + O(u^4) \ .
\label{lam6rtaylor}
\eeq
Hence, 
\beq
lim_{u \to 0} \frac{\lambda_{Z,sq2FP,j,u}}{\lambda_{Z,sq2FP,3,u}} = 
lim_{v \to \infty} \frac{\lambda_{Z,sq2FP,j}}{\lambda_{Z,sq2FP,3}} = 
\cases{ 1 & for $j=5$ \cr
        0 & for $j=1,2,4,6$ \cr } \ .
\label{lamratt0}
\eeq 
In this sense, one can define an effective, reduced, $N_{Z,G,\lambda}$,
and this has the value 2, somewhat analogous to the value
$N_{P,G,\lambda}=4$ for the $T=0$ limit of the Potts antiferromagnet
(\ref{pgsum}); however, a difference is that when one
reinserts the prefactor in (\ref{zu}) to get back the actual partition
function, all of the six $\lambda_{Z,G,j}$'s do contribute. 

\section{Coefficients for Other Strip Graphs}

\subsection{Strips with Torus or Klein Bottle Boundary Conditions}

Like the cyclic and M\"obius strips, the strips with torus or Klein bottle
boundary conditions have periodic or twisted periodic longitudinal boundary
conditions, respectively.  We first give two theorems for the sum of the
coefficients for these types of strip graphs (which are the analogues of
the theorems yielding eqs. (\ref{ncrelcyc}) and (\ref{csummb}) for the cyclic
and M\"obius strips).  We have

\begin{flushleft}

Theorem 10. \ The sum of the coefficients in (\ref{cpsum}) for a strip of the 
square or triangular lattice of width $L_y$ and 
$(PBC_y,PBC_x)=$ torus boundary conditions is (independent of $L_x$) 
\beq
C_{P,L_y,torus} = \sum_{j=1}^{N_{P,L_y,torus,\lambda}} c_{P,L_y,torus,j} = 
P(C_{L_y},q)
\label{cpsumtorus}
\eeq
where the chromatic polynomial of the circuit graph $C_n$ is 
\beq
P(C_n,q)=(q-1)^n+(q-1)(-1)^n \ . 
\label{pcn}
\eeq

\vspace{4mm}

Proof. \  This follows by noting that the transverse slice of the strip graphs
of the square and triangular lattices of width $L_y$ is the circuit graph 
$C_{L_y}$, and the number of ways of proper colorings of this graph is given by
the chromatic polynomial in (\ref{pcn}).  $\Box$ 

\vspace{4mm}

Theorem 11.  \ The sum of the coefficients in (\ref{cpsum}) for a strip of the
square or triangular lattice of width $L_y$ is (independent of $L_x$) and 
$(PBC_y,TPBC_x)=$ Klein bottle (KB) boundary conditions is 

\beq
C_{P,L_y,KB} = \sum_{j=1}^{N_{P,L_y,KB,\lambda}} c_{P,L_y,KB,j} = 0 
\label{cpsumklein}
\eeq

\vspace{4mm}

Proof. \ This follows by observing that this sum corresponds to the proper
coloring of an effective graph (defined as in our study of M\"obius strips) 
formed from the transverse slice of the strip, and in this case the effective
graph always involves at least one loop, i.e. an edge that connects a vertex to
itself (specifically, the effective graph involves one loop if $L_y$ is odd 
and two loops if $L_y$ is even).  But the chromatic polynomial of a graph with
a loop vanishes identically.  $\Box$ 

\vspace{4mm}

\end{flushleft}

Theorems 10 and 11 generalize the observations made for these sums of
coefficients for the exact solutions of the chromatic polynomials of the 
$L_y=3$ torus and Klein bottle graphs of the square lattice in \cite{tk} and
\cite{t} to arbitrary $L_y$ values where these graphs are defined, i.e. $L_y
\ge 3$. 

On the basis of the exact solutions for the chromatic polynomials for the width
$L_y=3$ strips of the square \cite{tk} and triangular \cite{t} lattices with
$(PBC_y,PBC_x)=$ torus and $(PBC_y,TPBC_x)=$ Klein bottle boundary conditions,
we can make several further remarks.  First, since it was found that
$N_{P,G,\lambda}=8$ for $G=sq(3 \times L_x,PBC_y,PBC_x)$ but the different
value, $N_{P,G,\lambda}=11$ for $tri(3 \times L_x,PBC_y, PBC_x)$, these strips
behave fundamentally differently than the cyclic and M\"obius strips of these
two lattices, for which the value of $N_{P,G,\lambda}$ was the same for a
given
$L_y$.  Secondly, the exact solutions in \cite{tk,t} show that the coefficients
that enter in the chromatic polynomial for the $L_y=3$ strips of the square and
triangular lattice with torus or Klein bottle boundary conditions are
polynomials in $q$ but are not of the form
$c^{(d)}=U_{2d}(\frac{\sqrt{q}}{2})$.  Indeed, in contrast to the situation for
cyclic and M\"obius square-lattice strips and cyclic triangular-lattice strips,
the coefficient of degree $d$ is not unique, i.e., there can be more than one
type of coefficient of a given degree $d$ in $q$.  For example, for the $L_y=3$
square-lattice strip with torus boundary conditions, the coefficients that are
quadratic functions of $q$ are $q(q-3)$, $(1/2)q(q-3)$, $(q-1)(q-2)$, and
$(1/2)(q-1)(q-2)$.  Another difference, again illustrated by the chromatic
polynomial for the $L_y=3$ strips of the square and triangular lattice is that
the coefficients can have zeros outside the range $0 < q < 4$, in contrast to
the $c^{(d)}$. For example, one term in the chromatic polynomial for the
$L_y=3$ strip with toroidal boundary conditions is $(q^3-6q^2+8q-1)(-1)^{L_x}$
for the square lattice \cite{tk} and $(1/3)(q-1)(q^2-5q+3)(-2)^{L_x}$ for the
triangular lattice \cite{t};
the first coefficient has a zero at $q \simeq 4.11$ and the other coefficient
has a zero at $q \simeq 4.30$.  Given that the exact solutions for the
chromatic polynomials on torus graphs in \cite{tk,t} show that the
coefficients, which are multiplicities of eigenvalues of the coloring matrix,
are not of the form $c^{(d)}$ for strip graphs of the square and triangular
lattices, it is not clear to us how to relate this with the remark in
\cite{baxter} that certain related eigenvalue multiplicities $d_r$ for torus
boundary conditions are of the
form $\sin(r\theta/2)/\sin(\theta/p)$ where $p$ is 1 (2) if $r$ is even (odd)
and $\theta$ satisfies (\ref{qtrel}).

The exact solutions for the chromatic polynomial in \cite{tk} for the square
lattice and in \cite{t} for the triangular lattice showed that if one starts
with torus boundary conditions, cuts the tube graph and reglues the ends with
reversed orientation to form Klein bottle boundary conditions, the number of
$\lambda_{P,G,j}$'s does change (is reduced):
\beq
N_{P,\Lambda,L_y \times L_x,PBC_y,PBC_x} \ne 
N_{P,\Lambda,L_y \times L_x,PBC_y,TPBC_x} \ .
\label{ntormb}
\eeq
Specifically, it was found that \cite{tk,bcc,t} while 
\beq
N_{P,sq,3 \times L_x,PBC_y,PBC_x,\lambda}=8 \ , \quad 
N_{P,tri,3 \times L_x,PBC_y,PBC_x,\lambda}=11
\label{nptorussqtri}
\eeq
in contrast, 
\beq
N_{P,sq,3 \times L_x,PBC_y,TPBC_x,\lambda}=
N_{P,tri,3 \times L_x,PBC_y,PBC_x,\lambda}=5 \ .
\label{nsqtriklein}
\eeq 
Thus, at least for this width, if one uses Klein bottle boundary
conditions, then the value of $N_{P,G,\lambda}$ is the same, namely five, for
both square and triangular strips.  As is evident from the exact solutions
\cite{tk,t}, the five coefficients are the same, although the
$\lambda_{P,G,j}$'s are different.  These coefficients are 
\beq 
\{ 1, \ q-1, \ -(q-1), \ -\frac{(q-1)(q-2)}{2}, \ \frac{q(q-3)}{2} \}
\label{coefftorus}
\eeq
Since the coloring matrix applies directly for the lattice strip
graphs with torus boundary conditions, eq. (\ref{ttracezero}) applies for the
chromatic polynomials of these graphs.

\subsection{Strips with Free Longitudinal Boundary Conditions}

Previously, exact solutions for chromatic polynomials on strip graphs of
regular lattices with $(FBC_y,FBC_x)=$ open boundary conditions were given in
\cite{strip,hs}; in these cases, as discussed before, the coefficients
$c_{P,G,j}$ are not, in general, polynomials in $q$.  This can be seen
immediately from eqs. (2.14) or (2.19) of \cite{hs}. An analogous statement
holds for the full Potts model partition functions on these open strips
\cite{bcc,a,ta}.  This property can be understood as a consequence of the fact
that the chromatic polynomial is not the trace of an $m$'th power of the
coloring matrix for these strips, but rather is given by 
\beq
P(G,q) = \langle c | {\cal T}^m | c^\prime \rangle
\label{pgopen}
\eeq
where $c$ and $c^\prime$ denote colorings of the ends of the strip.  A similar
statement holds for the full Potts partition function $Z(G,q,v)$ in terms of
the corresponding matrix ${\cal T}_Z$.  

As discussed before \cite{w2d,pg,wcy,pm,bcc,t}, for a strip graph of some
lattice with given transverse boundary conditions, the term $\lambda_{P,G,j}$
that is dominant in the physical region of $q$ (including sufficiently positive
integers and denoted as region $R_1$ in our previous work \cite{w}) is
independent of the longitudinal boundary conditions. (The same applies for the
$\lambda_{Z,G,j}$ in $Z(G,q,v)$ that is dominant in the physical paramagnetic
phase \cite{bcc,a,ta}.) In particular, this means that the $\lambda_{P,G,j}$
that is dominant in region $R_1$ is the same for the cyclic (or M\"obius) and
open strip graphs of a given lattice type $\Lambda$, thereby establishing a
certain connection between these cyclic (or M\"obius) and open strip graphs.
Similarly, it means that the $\lambda_{P,G,j}$ that is dominant in region $R_1$
is the same for the torus and Klein bottle strips of this lattice $\Lambda$ on
the one hand, and the cylindrical strip graphs on the other.  Now it was also
shown \cite{bcc} that for graphs with periodic or twisted periodic boundary
conditions, the coefficient of this dominant term is unity.  If this term is a
root $r_{\ell,s}$ of an (irreducible) algebraic equation of degree $d_\ell$,
then the theorem on symmetric functions of roots of an algebraic equation
implies that for the cyclic strips, these roots $r_{\ell,s}$ appear in $P(G,q)$
in the form of the sum $\sum_{s=1}^{d_\ell} r_{\ell,s}^{L_x}$, i.e., they all
have the coefficient unity \cite{pm}.  This sets an upper bound on
$d_\ell(\Lambda,L_y)$, namely 
\beq 
n_P(L_y,0) \ge d_\ell(\Lambda,L_y)
\label{npineq}
\eeq
where we have included the dependent of $d_\ell$ on the lattice type $\Lambda$
and the strip width, and we recall from eq. (\ref{lameqcycmb}) that 
$d_\ell(\Lambda,L_y)$ is the same for the cyclic and M\"obius strips of a given
lattice $\Lambda$.  For cyclic and M\"obius strips of the square lattice, 
$d_\ell(sq,1)=1$, $d_\ell(sq,2)=1$ \cite{bds}, $d_\ell(sq,3)=2$  \cite{wcy},
and $d_\ell(sq,4)=3$ \cite{s4}, while for the cyclic and M\"obius strips of the
triangular lattice, $d_\ell(tri,2)=1$ \cite{wcy} and $d_\ell(tri,3)=2$,
$d_\ell(tri,4)=4$ \cite{t}.  
This information is summarized in tables given in \cite{t,s4} along
with related results. Our calculations above yield, for the relevant values of
$n_P(L_y,0)$ (see Table \ref{npctablecyc}): 
$n_P(1,0)=n_P(2,0)=1$, $n_P(3,0)=2$,
and $n_P(4,0)=4$.  Hence, evidently, the inequality (\ref{npineq}) is realized
as an equality for the triangular strips and the square strips for $L_y=1,2,3$,
but as a strict inequality for the $L_y=4$ square strips.  Further, for the
cases that we studied, we found that 
\beq
d_\ell(\Lambda,L_y)_{cyc.,Mb}=d_\ell(\Lambda,L_y)_{open}
\label{dcycopen}
\eeq
where $d_\ell(\Lambda,L_y)_{open} \equiv d_\ell(\Lambda,FBC_y,FBC_x,L_y)$. 
When we obtained the results in the present paper for $n_P(L_y,d)$, 
we noticed the interesting connection that for the cyclic and
M\"obius strips of the triangular lattice, for which (\ref{npineq}) holds as 
an equality, it is also true that one has the fourfold equality 
\beq
n_P(L_y,0) = d_\ell(tri,L_y)_{cyc., \ Mb.} = d_\ell(tri,L_y)_{open}=
N_{P,tri(L_y),open,\lambda}
\label{nply0nptotopen}
\eeq
i.e., the number of $\lambda_{P,G,j}$'s with coefficient $c^{(0)}=1$ in the
chromatic polynomial for the cyclic or M\"obius strips of the triangular 
lattice, which was equal to the degree of the algebraic equation one of whose
roots was the dominant $\lambda_{P,G,j}$ in $R_1$, was also equal to 
the total number of terms in the chromatic polynomial 
for the open strip of the triangular lattice with the same width.  We confirmed
eq. (\ref{nply0nptotopen}) for widths $L_y=2,3,4$, based on the exact 
solutions in \cite{strip,wcy,t}.   In terms of the 
generating function formalism of \cite{strip,strip2,hs}, the denominator 
${\cal D}$ of the generating function, as a polynomial in the auxiliary 
variable, has the maximal degree, $n_P(L_y,0)$, allowed by (\ref{npineq}) for
these open strips of the triangular lattice but does not, in general, for the 
open strips of the square lattice.  As we noted in \cite{t,s4}, this 
difference in behavior is related to the fact that the M\"obius strips of the 
square and triangular lattices are different; in particular, the M\"obius strip
of the triangular lattice involves algebraic nonpolynomial coefficients. 
Further, even for the strips of the square
lattice, where the first two terms of (\ref{nply0nptotopen}) are not, in
general, equal (as our solutions for chromatic polynomials for $L_y=4$ showed),
we still found 
\beq
d_\ell(sq,L_y)_{cyc.,Mb.}=N_{P,sq(L_y),open,\lambda} \ .
\label{nply0nptotsqopen}
\eeq
That is, we find that the equality (\ref{nply0nptotsqopen}) holds for cyclic 
or M\"obius and open strips of the square lattice for $1 \le L_y \le 4$, as 
well as for the corresponding strips of the triangular lattice with $2 \le L_y
\le 4$; in particular, it holds even when $n_P(L_y,0)$ is not equal to (is
greater than) $d_\ell(\Lambda,L_y)$. For reference, for open strips of the 
triangular lattice with $L_y \ge 2$, one can infer from calculations in 
\cite{baxter} that $N_{Z,tri(L_y),open,\lambda}=C_{L_y}$; furthermore, 
$N_{P,tri(L_y),open,\lambda}=M_{L_y-1}$ \cite{ss}. 
Another generalization would be that for triangular strips, the value of the
degree in eq. (\ref{dcycopen}) satisfies 
$d_\ell(tri,L_y)_{cyc.,Mb}=d_\ell(tri,L_y)_{open}=M_{L_y-1}$.
This would require that, in terms of the generating function \cite{strip}, the
denominator ${\cal D}$ does not factorize.  

Previously, coloring matrix methods have been applied to the honeycomb ($hc$)
lattice to derive rigorous bounds on $W$ \cite{ww,wn}.  In the present context
of strips, we observe that, in addition to the elementary result
$N_{P,hc,L_y=2,open,\lambda}=d_\ell(hc,L_y=2)_{open}=1$, our exact solutions
have yielded the results 
\beq 
N_{P,hc,L_y=3,open\lambda}=d_\ell(hc,L_y=3)_{open}=3
\label{dellhc3open}
\eeq
 from \cite{strip} and
\beq
N_{P,hc,L_y=2,cyc.,Mb.,\lambda}=d_\ell(hc,L_y=2)_{cyc.,Mb}=1
\label{dellhc2cyc}
\eeq
 from \cite{pg}.  In view of eq. (\ref{dellhc3open}), for this strip of the
 honeycomb lattice, $n_P(L_y=3,0) \ne d_\ell(hc,L_y=3)_{open}$, in
 contrast to the case with the square and triangular lattice strips.

\subsection{Strips with Cylindrical Boundary Conditions}

Exact solutions for chromatic polynomials for cylindrical strips, i.e. with
boundary conditions of the form $(PBC_y,FBC_x)$ of 
the square and triangular lattices were presented in \cite{strip2,w2d}, and,
again, the coefficients are, in general, not polynomials in $q$, for the same
reason as in the case of the open strips.  And again, the same statement holds
for the full Potts model partition function.  (Recent calculations of chromatic
polynomials for wider strips with open and cylindrical boundary conditions
include \cite{t,s4,ss}.) 

\section{Some General Geometric Identities}

In this section we present some useful identities between Potts model partition
functions on different types of lattice strips that follow from basic
geometrical considerations.  We have applied these as checks in our previous
calculations. Consider a family of strip graphs of the square or triangular
lattice with fixed width $L_y$ and arbitrary length $L_x$ with some set of
longitudinal ($x$) and transverse ($y$) boundary conditions, denoted as
$BC=(BC_y,BC_x)$. By the transfer matrix argument given in \cite{a}, it follows
that the Potts model partition function for this strip is of the form 
\beq
Z(\Lambda, L_y \times L_x, BC,q,v) = \sum_{j=1}^{N_{Z,\Lambda,L_y,BC}}
c_{Z,\Lambda,L_y,BC,j}(\lambda_{Z,\Lambda,L_y,BC,j})^{L_x} \ .
\label{zff1}
\eeq
We recall \cite{a} that the total number of terms, $N_{Z,\Lambda,L_y,BC}$, the 
coefficients $c_{Z,\Lambda,L_y,BC,j}$, and the 
$\lambda_{Z,\Lambda,L_y,BC,j}$'s are independent of $L_x$.  In our previous
works, we have studied the $L_x \to \infty$ behavior of these Potts model
partition functions.  However, clearly, for fixed $L_y$ and $L_x$, by switching
the $x$ and $y$ axes, one can equivalently view the strip as the length--$L_y$
member of a family of strip graphs with fixed width $L_x$ and length
$L_y$.\footnote{\footnotesize{
Note that for other lattices such simple relations do not, in general,
hold.  For example, if one considers strips of the honeycomb lattice,
constructed as a brick lattice, then rotating a open horizontal $L_y \times
L_x$ strip of bricks by $90^\circ$, one does not get an equivalent $L_x \times
L_y$ strip of horizontally oriented bricks.}} This yields a number of useful
identities.  We proceed to describe these 

\subsection{Strips with $(FBC_y,FBC_x)$}

For strips of the square and triangular lattices with $(FBC_y,FBC_x)$ (open)
boundary conditions, the interchange of the $x$ and $y$ axes leaves the
boundary conditions the same, and one gets the following identity (where we
write $F_yF_x \equiv (FBC_y,FBC_x)$ to save space)
\beq
\sum_{j=1}^{N_{Z,\Lambda,L_y,F_y,F_x}}
c_{Z,\Lambda,L_y,F_y,F_x,j}(\lambda_{Z,\Lambda,L_y,F_y,F_x,j})^{L_x}=
\sum_{j=1}^{N_{Z,\Lambda,L_x,F_x,F_y}}
c_{Z,\Lambda,L_x,F_x,F_y,j}(\lambda_{Z,\Lambda,L_x,F_x,F_y,j})^{L_y} \ .
\label{zff}
\eeq 
Taking $T=0$ for the Potts
antiferromagnet, one obtains the corresponding identity for chromatic
polynomials 
\beq 
\sum_{j=1}^{N_{P,\Lambda,L_y,F_y,F_x}}
c_{P,\Lambda,L_y,F_y,F_x,j}(\lambda_{P,\Lambda,L_y,F_y,F_x,j})^{L_x}=
\sum_{j=1}^{N_{P,\Lambda,L_x,F_y,F_x}}
c_{P,\Lambda,L_x,F_x,F_y,j}(\lambda_{P,\Lambda,L_x,F_x,F_y,j})^{L_y} \ .
\label{pff}
\eeq
Using the results in \cite{strip}, one can construct various illustrations of
this.  One of the simplest is to take the open strip of the square lattice with
width $L_y=2$, for which $N_{P,sq,2,FBC_y,FBC_x}=1$ and 
$P(sq,2 \times L_x,FBC_y,FBC_x,q)=q(q-1)(q^2-3q+3)^{L_x-1}$.  For $L_x=3$, this
chromatic polynomial must be equal to that for the width $L_y=3$ open strip
(which has $N_{P,sq,3,FBC_y,FBC_x}=2$) with length $L_x=2$, for which (using
\cite{strip} and eq. (2.15) of \cite{hs}) for general $L_x$, 
\beqs
& & P(sq,3 \times L_x,FBC_y,FBC_y,q)= 
\frac{1}{(\lambda_{sq3FF,1}-\lambda_{sq3FF,2})} \times \cr\cr
& & 
\Bigl [ (A_{sq3FF,0} \lambda_{sq3FF,1} + A_{sq3FF,1})
(\lambda_{sq3FF,1})^{L_x-2}
-(A_{sq3FF,0} \lambda_{sq3FF,2}+A_{sq3FF,1})(\lambda_{sq3FF,2})^{L_x-2} \Bigr ]
\cr\cr
& & 
\label{psq3ff}
\eeqs
where 
\beq
A_{sq3FF,0}=q(q-1)(q^2-3q+3)^2 
\label{asq3ff}
\eeq
\beq
A_{sq3FF,1}=-q(q-1)^3(q^3-6q^2+13q-11)
\label{asq3ff1}
\eeq
and
\beq
\lambda_{sq3FF,(1,2)}=\frac{1}{2}\biggl [ (q-2)(q^2-3q+5) \pm 
\Bigl [(q^2-5q+7)(q^4-5q^3+11q^2-12q+8) \Bigr ]^{1/2} \biggr ] \ .
\label{lamsq3ff}
\eeq

\subsection{Cyclic and Cylindrical Strips}

We have denoted strips of fixed width $L_y$ and arbitrary length $L_x$ vertices
with $(FBC_y,PBC_x)$ and $(PBC_y,FBC_x)$ boundary conditions as cyclic and
cylindrical, respectively.  Topologically, these are the same; the distinction
was made because we were, in particular, interested in the large $L_x \to
\infty$ limit, which, for cyclic strips means that the length of a circuit goes
to infinity, while for the cylindrical strips, the length of the circumference
of the cylinder is fixed while its length goes to infinity.  Clearly, a cyclic
strip of the square or triangular lattice with a fixed width $L_y$ and length
$L_x$ is the same as a cylindrical strip of the given lattice with width $L_x$
and length $L_y$.  Letting $\Lambda$ denote the type of lattice, square or
triangular as before, and writing 
\beq
Z(\Lambda, L_y \times L_x, FBC_y,PBC_x)=\sum_{j=1}^{N_{Z,\Lambda,L_y,F_y,P_x}}
c_{Z,\Lambda,L_y,F_y,P_x,j}(\lambda_{Z,\Lambda,L_y,F_y,P_x,j})^{L_x}
\label{zff2}
\eeq
we have the identity 
\beq
 \sum_{j=1}^{N_{Z,\Lambda,L_y,F_y,P_x}}
c_{Z,\Lambda,L_y,F_y,P_x,j}(\lambda_{Z,\Lambda,L_y,F_y,P_x,j})^{L_x}=
 \sum_{j=1}^{N_{Z,\Lambda,L_x,P_y,F_x}}
c_{Z,\Lambda,L_x,P_y,F_x,j}(\lambda_{Z,\Lambda,L_x,P_y,F_x,j})^{L_y} \ .
\label{zfppf}
\eeq
Again, for the special case of the $T=0$ Potts antiferromagnet, we have the
resultant identity for the chromatic polynomial
\beq
 \sum_{j=1}^{N_{P,\Lambda,L_y,F_y,P_x}}
c_{P,\Lambda,L_y,F_y,P_x,j}(\lambda_{P,\Lambda,L_y,F_y,P_x,j})^{L_x}=
 \sum_{j=1}^{N_{P,\Lambda,L_x,P_y,F_x}}
c_{P,\Lambda,L_x,P_y,F_x,j}(\lambda_{P,\Lambda,L_x,P_y,F_x,j})^{L_y} \ .
\label{pfppf}
\eeq
Perhaps the simplest illustration of this general identity is the relation
between the chromatic polynomial for the width $L_y=2$ cyclic strip of the 
square lattice and the width $L_y=3$ cylindrical strip of this lattice.  For
the $L_y=2$ cyclic strip, one has $N_{P,sq,2 \times L_x,F_y,P_x}=4$ and 
\beq
P(sq,2 \times L_x,FBC_y,PBC_x,q)=(q^2-3q+1)+ (q-1)\Bigl [ (3-q)^{L_x}+
(1-q)^{L_x} \Bigr ] + (q^2-3q+3)^{L_x} \ .
\label{plad}
\eeq
For the $L_y=3$ cylindrical strip, one has 
\beq
P(sq,3 \times L_x,PBC_y,FBC_x,q)=q(q-1)(q-2)(q^3-6q^2+14q-13)^{L_x-1} \ .
\label{pcyl}
\eeq
Thus, the identity (\ref{pfppf}) yields 
\beqs
& & (q^2-3q+1)+ (q-1)\Bigl [ (3-q)^3 + (1-q)^3 \Bigr ] + (q^2-3q+3)^3 = \cr\cr
& & q(q-1)(q-2)(q^3-6q^2+14q-13) \ .
\label{pcyccyl}
\eeqs
While this illustration is quite simple, more complicated cases involve
identities between powers of algebraic roots of different types. 

\subsection{Strips with Torus Boundary Conditions} 

For a strip graph of the square or triangular lattice with $(PBC_y,PBC_x)
\equiv (P_y,P_x)$ (torus) boundary conditions, the identities read 
\beq
 \sum_{j=1}^{N_{Z,\Lambda,L_y,P_y,P_x}}
c_{Z,\Lambda,L_y,P_y,P_x,j}(\lambda_{Z,\Lambda,L_y,P_y,P_x,j})^{L_x}=
 \sum_{j=1}^{N_{Z,\Lambda,L_x,P_x,P_y}}
c_{Z,\Lambda,L_x,P_x,P_y,j}(\lambda_{Z,\Lambda,L_x,P_x,P_y,j})^{L_y}
\label{zpppp}
\eeq

\beq
 \sum_{j=1}^{N_{P,\Lambda,L_y,P_y,P_x}}
c_{P,\Lambda,L_y,P_y,P_x,j}(\lambda_{P,\Lambda,L_y,P_y,P_x,j})^{L_x}=
 \sum_{j=1}^{N_{P,\Lambda,L_x,P_x,P_y}}
c_{P,\Lambda,L_x,P_x,P_y,j}(\lambda_{P,\Lambda,L_x,P_x,P_y,j})^{L_y}
\label{ppppp}
\eeq
The simplest illustration of this is for the family of strips of the square
lattice with width $L_y=2$, for which the torus degenerates to a cyclic strip,
and one has $P(sq,2 \times L_x,PBC_y,PBC_x,q)=P(sq,2 \times
L_x,FBC_y,PBC_x,q)$, given in (\ref{plad}).  Let us next consider the strip of
the square lattice with width $L_y=3$ and length $L_x=2$. 
This graph is related by the identity to two other strip graphs: (i) the cyclic
strip of the square lattice with $L_y=2$ and $L_x=3$ (see eq. (\ref{plad})) and
(ii) the equivalent cylindrical strip of the square lattice with $L_y=3$ 
and $L_x=2$ (see eq. (\ref{pcyl})).  Indeed, when one evaluates the exact
solution given in \cite{tk} for $L_x=2$, one finds that it is equal to the
right-hand side of eq. (\ref{pcyccyl}).

\section{Conclusions}

In this paper, using a general formula, in terms of Chebyshev polynomials of
the second kind, for the coefficients that occur in the $q$-state Potts model
partition function on cyclic strips of the square and triangular lattices and
on M\"obius strips of the square lattice, together with general formulas for
sums of these coefficients, we have determined the number, $n_Z(L_y,d)$, of
$\lambda_{Z,G,j}$'s with coefficient $c^{(d)}$ in $Z(G,q,v)$ and the total
number, $N_{Z,G,\lambda}$, for these strips.  Results are also given for the
analogous numbers $n_P(L_y,d)$ and $N_{P,L_y,\lambda}$ for the zero-temperature
Potts antiferromagnet partition functions, i.e., chromatic polynomials for
these strips.  The results for the total numbers, $N_{Z,L_y,\lambda}$ and
$N_{P,L_y,\lambda}$, apply for both cyclic and M\"obius strips of both the
square and triangular lattices.  Among other connections, we found that
$n_P(L_y,0)=n_P(L_y-1,1)=M_{L_y-1}$, the Motzkin number; $n_Z(L_y,0)=C_{L_y}$,
the Catalan number; the exact expression for $N_{P,L_y,\lambda}$ in
eq. (\ref{nptotform}); and the relations $N_{P,L_y,\lambda}=2N_{DA,sq,L_y}$;
and $N_{Z,L_y,\lambda}= 2N_{DA,tri,L_y}$, where $N_{DA,\Lambda,n}$ denotes the
number of directed lattice animals on the lattice $\Lambda$.  We also found the
asymptotic growths $N_{Z,L_y,\lambda} \sim L_y^{-1/2} \ 4^{L_y}$ and
$N_{P,L_y,\lambda} \sim L_y^{-1/2} \ 3^{L_y}$ as $L_y \to \infty$.  Some
commens about other lattice strips were made.  In addition, we presented some
useful general geometric identities for Potts model partition functions.

\vspace{10mm}

Acknowledgment: We thank Lee-Peng Teo for helpful discussions. 
The research of R. S. was supported in part by the NSF
grant PHY-9722101 and at Brookhaven by the DOE contract
DE-AC02-98CH10886.\footnote{\footnotesize{Accordingly, the U.S. government
retains a non-exclusive royalty-free license to publish or reproduce the
published form of this contribution or to allow others to do so for
U.S. government purposes.}}

\vspace{6mm}

\section{Appendix}

\subsection{Connection Between Potts Model Partition Function and Tutte
Polynomial}

The Potts model partition function $Z(G,q,v)$ is related to the Tutte
polynomial $T(G,x,y)$ as follows.
The graph $G$ has vertex set $V$ and edge set $E$,
denoted $G=(V,E)$.  A spanning subgraph $G^\prime$ is defined as a subgraph
that has the same vertex set and a subset of the edge set:
$G^\prime=(V,E^\prime)$ with $E^\prime \subseteq E$.  The Tutte polynomial
of $G$, $T(G,x,y)$, is then given by \cite{tutte1}-\cite{tutte3}
\beq
T(G,x,y)=\sum_{G^\prime \subseteq G} (x-1)^{k(G^\prime)-k(G)}
(y-1)^{c(G^\prime)}
\label{tuttepol}
\eeq
where $k(G^\prime)$, $e(G^\prime)$, and $n(G^\prime)=n(G)$ denote the number
of components, edges, and vertices of $G^\prime$, and
\beq
c(G^\prime) = e(G^\prime)+k(G^\prime)-n(G^\prime)
\label{ceq}
\eeq
is the number of independent circuits in $G^\prime$ (sometimes called the
co-rank of $G^\prime$).   Note that the first factor can also be written as
$(x-1)^{r(G)-r(G^\prime)}$, where
\beq
r(G) = n(G)-k(G)
\label{rank}
\eeq
is called the rank of $G$.  The graphs $G$ that we consider here are
connected, so that $k(G)=1$.  Now let
\beq
x=1+\frac{q}{v}
\label{xdef}
\eeq
and
\beq
y=a=v+1
\label{ydef}
\eeq
so that $q=(x-1)(y-1)=(x-1)v$. Then
\bigskip
\beq
Z(G,q,v)=(x-1)^{k(G)}(y-1)^{n(G)}T(G,x,y) \ .
\label{ztutte}
\eeq
There is also a connection with the Whitney rank polynomial,
$R(G,\xi,\eta)$, defined as \cite{whit,bbook}
\beq
R(G,\xi,\eta)=\sum_{G^\prime \subseteq G}\xi^{r(G^\prime)}\eta^{c(G^\prime)}
\label{whitney}
\eeq
where the sum is again over spanning subgraphs $G^\prime$ of $G$. Then
\beq
T(G,x,y)=(x-1)^{r(G)}R(G,\xi=(x-1)^{-1},\eta=y-1)
\label{tr}
\eeq
and
\beq
Z(G,q,v)=q^{n(G)} R(G,\xi=\frac{v}{q},\eta=v) \ .
\label{zwhit}
\eeq
Note that the chromatic polynomial is a special case of the Tutte polynomial:
\beq
P(G,q)=q^{k(G)}(-1)^{k(G)+n(G)}T(G,x=1-q,y=0)
\label{tprel}
\eeq
(recall eq. (\ref{zp})).

For a recursive family of graphs, such as the strip graphs considered in this
paper, comprised of $m$ repetitions of a basic subgraph, the Tutte polynomial
has the form \cite{a}
\beq
T(G,x,y) = \sum_{j=1}^{N_{T,G,\lambda}} c_{T,G,j}(\lambda_{T,G,j})^m
\label{tgsum}
\eeq
where, from (\ref{ztutte}), one has the relation 
\beq
\lambda_{T,G,j}= v^{-L_y}\lambda_{Z,G,j}
\label{lamtlamz}
\eeq
so that
\beq
N_{T,G,\lambda}=N_{Z,G,\lambda} \ . 
\label{nttotnztoteq}
\eeq
It is convenient to extract a common factor from the coefficients:
\beq
c_{T,G,j} \equiv \frac{\bar c_{T,G,j}}{x-1} \ .
\label{cbar}
\eeq
Of course, although the individual terms contributing
to the Tutte polynomial are thus rational functions of $x$ rather than
polynomials in $x$, the full Tutte polynomial is a polynomial
in both $x$ and $y$.  
Given the relation (\ref{ztutte}), if one defines $n_T(L_y,d)$
as the number of terms $\lambda_{T,L_y,j}$ in $T(L_y,q,v)$ that have as their
reduced coefficient $\bar c_{T,L_y,j}=c^{(d)}$, then these are the same
numbers:
\beq
n_T(L_y,d)=n_Z(L_y,d) \ . 
\label{ntnzeq}
\eeq
Thus, Tables \ref{ntctablecyc} and \ref{ntctablemb} apply equally to the
structure of the Tutte polynomials for the cyclic and M\"obius strips of the
square and triangular lattices. 

For a given graph $G=(V,E)$, at certain special values of the arguments $x$ and
$y$, the Tutte polynomial $T(G,x,y)$ yields quantities of basic graph-theoretic
interest \cite{tutte3}-\cite{welsh}.  We recall some definitions: a spanning
subgraph was defined at the beginning of the paper; a tree is a
connected graph with no cycles; a forest is a graph containing one or
more trees; and a spanning tree is a spanning subgraph that is a tree.  We
recall that the graphs $G$ that we consider are connected.  Then the number
of spanning trees of $G$, $N_{ST}(G)$, is
\beq
N_{ST}(G)=T(G,1,1) \ ,
\label{t11}
\eeq
the number of spanning forests of $G$, $N_{SF}(G)$, is
\beq
N_{SF}(G)=T(G,2,1) \ ,
\label{t21}
\eeq
the number of connected spanning subgraphs of $G$, $N_{CSSG}(G)$, is
\beq
N_{CSSG}(G)=T(G,1,2) \ ,
\label{T12}
\eeq
and the number of spanning subgraphs of $G$, $N_{SSG}(G)$, is
\beq
N_{SSG}(G)=T(G,2,2) \ .
\label{t22}
\eeq
These connections have been used in \cite{wu77},\cite{ws}. 

\subsection{Determinants of Coloring Matrices}

In this section we list the determinants of coloring matrices for some lattice
strips. For the chromatic polynomial this is 
\beq 
\det T_P(G) = \prod_{j=1}^{\cal N} \lambda_{P,G,j} = \prod_{j=1}^{N_{P,G,
\lambda}} (\lambda_{P,G,j})^{c_{P,G,j}} \ .
\label{detformp}
\eeq
For the Potts model partition function the analogous determinant is
\beq
\det T_Z(G) = \prod_{j=1}^{\cal N_Z} \lambda_{Z,G,j} =
\prod_{j=1}^{N_{Z,G,
\lambda}} (\lambda_{Z,G,j})^{c_{Z,G,j}} \ .
\label{detformz}
\eeq

Let us define the shorthand notation
\beq
D_P(G)=\det T_P(G) \ , \quad D_Z(G)=\det T_Z(G) \ .
\label{dnotation}
\eeq
For the products of eigenvalues contributing to (\ref{pgsum}) for the chromatic
polynomials and (\ref{zgsum}) for the Potts model partition function for cyclic
and M\"obius strips of the square lattice we find 
\beq
D_P(sq,L_y=1,FBC_y,PBC_x)=D_P(\{C\},q)=(-1)^{q-1}(q-1)
\label{dsqly1p}
\eeq
(where $\{C\}$ refers to the circuit graph)
For the $L_y=2$ cyclic and M\"obius strips of the square lattice, from 
\cite{bds}, 
\beq
D_P(sq,L_y=2,FBC_y,PBC_x)=\Bigl [ (3-q)(1-q) \Bigr ]^{q-1}(q^2-3q+3)
\label{dsqly2p}
\eeq
and 
\beq
D_P(sq,L_y=2,FBC_y,TPBC_x)= (3-q)^{q-1}(1-q)^{-(q-1)}(q^2-3q+3) \ . 
\label{dsqly2mbp}
\eeq
For the $L_y=3$ cyclic and M\"obius strips of the square lattice, from
\cite{wcy}, 
\beqs
& & D_P(sq,L_y=3,FBC_y,PBC_x) = (-1)^{(q-2)(q^2-3q+1)}(q-1)^{(q-1)^2}
(q-2)^{q^2-q-1} \times \cr\cr
& & (q-4)^{q^2-3q+1}(q^3-6q^2+13q-11)(q^4-9q^3+29q^2-40q+22)^{q-1}
\label{dsqly3p}
\eeqs
and 
\beqs
& & D_P(sq,L_y=3,FBC_y,TPBC_x) = 
(-1)^{q^2-4q+2}(q-1)^{q-1}(q-2)^{-(2q-3)} \times 
\cr\cr
& & (q-4)^{-1}(q^3-6q^2+13q-11)(q^4-9q^3+29q^2-40q+22)^{q-1} \ . 
\label{dsqly3mbp}
\eeqs
The corresponding determinants for $L_y=4$ can be obtained from \cite{s4} but
are quite lengthy, so we do not list them here. 

For the $D_Z$ determinants we have
\beq
D_Z(sq,L_y=1,FBC_y,PBC_x)=D_Z(\{C\},q)=v^{q-1}(q+v)
\label{dsqly1z}
\eeq
and, from \cite{a}, 
\beq
D_Z(sq,L_y=2,FBC_y,PBC_x)=v^{2q(q-1)}(v+1)^q(v+q)^{2q}
\label{dsqly2z}
\eeq
and 
\beq
D_Z(sq,L_y=2,FBC_y,TPBC_x)=v^{2(q-1)}(v+1)^q(v+q)^2 \ . 
\label{dsqly2mbz}
\eeq 
Note that, except for the case of the circuit graph, $sq,L_y=1,PBC_x$, for
the value $v=-1$, where the Potts model partition function reduces to the
chromatic polynomial, $Z(G,q,v=-1)=P(G,q)$, some eigenvalues in the
product
contributing to $Z$ vanish, and hence this product vanishes.  If one extracts
these vanishing eigenvalues, then, of course, the rest yield the same product
as for $P(G,q)$.

For cyclic strips of the triangular lattice we have, from \cite{wcy}

\beq
D_P(tri,L_y=2,FBC_y,PBC_x)=(q-2)^{2q}
\label{dtrily2p}
\eeq
and from \cite{t,ta}, 
\beq
D_P(tri,L_y=3,FBC_y,PBC_x)=(-1)^{(q-2)(q^2-3q+1)}(q-2)^{q(2q-1)}(q-3)^{q(q-1)}
\label{dtrily3p}
\eeq

\beq
D_Z(tri,L_y=2,FBC_y,PBC_x)=v^{2q(q-1)}(v+1)^{2q}(v+q)^{2q} \ .
\label{dtrily2z}
\eeq
The determinant $D_P(tri,L_y=4,FBC_y,PBC_x)$ can be calculated from our exact
solution in \cite{t}; however, it is sufficiently lengthy that we do not
include it here.  The analogous determinants for the M\"obius strips of the
triangular lattice can also be calculated from the exact solutions that we have
given \cite{wcy,t}; however, they are more complicated, since some coefficients
are algebraic, rather than polynomial, functions of $q$.

For the strip graphs of the square and triangular lattices with
$(PBC_y,PBC_x)=$ torus and $(PBC_y,TPBC_x)=$ Klein bottle boundary conditions,
we have, from the exact solutions in \cite{tk} and \cite{t}, 
\beqs
& & D_P(sq,L_y=3,PBC_y,PBC_x)=(-1)^{q^3-6q^2+11q-4}(q-1)^{\frac{(q-1)(q-2)}{2}}
\times \cr\cr
& & 
(q-2)^{q^2+q-4}(q-4)^{(q-1)(q-2)}(q-5)^{\frac{q(q-3)}{2}}
(q^2-7q+13)^{q-1} \times \cr\cr
& & (q^3-6q^2+14q-13)
\label{dsqly3torus}
\eeqs
and 
\beqs
D_P(sq,L_y=3,PBC_y,TPBC_x) & = & (q-1)^{-\frac{(q-1)(q-2)}{2}}
(q-5)^{\frac{q(q-3)}{2}}(q^2-7q+13)^{q-1} \times \cr\cr
& & (q^3-6q^2+14q-13) \ . 
\label{dsqly3klein}
\eeqs

\vspace{8mm}

\subsection{Some Other Coloring Matrix Results for Lattice Strip Graphs} 

Since the trace of a coloring matrix vanishes (recall (\ref{ttracezero})),
the 
sum of eigenvalues, each multiplied by its multiplicity also vanishes, as
indicated in eq. (\ref{clamzero}), for the cyclic and torus strips, for which 
(\ref{pttrace}) applies directly.  Using coloring methods, we obtain the
following general formulas for sums for other types of strips.  These agree 
with our previous exact solutions in \cite{wcy,t,s4}. 

\beq
\sum_{j=1}^{N_{P,sq(L_y),Mb,\lambda}} c_{P,sq(L_y),Mb,j}
\lambda_{P,sq(L_y),Mb,j}= 
\cases{ q(q-1)(q^2-3q+3)^{\frac{L_y}{2}-1} & for even $L_y$ \cr
0 & for odd $L_y$ \cr }
\label{sqmblamsum}
\eeq

\beq
\sum_{j=1}^{N_{P,tri(L_y),Mb,\lambda}} c_{P,tri(L_y),Mb,j}
\lambda_{P,tri(L_y),Mb,j}=0 \quad \forall \ L_y \ge 2 \ . 
\label{trimblamsum}
\eeq
The equations also apply to the corresponding strips with Klein bottle (KB) 
boundary conditions, with $Mb \to KB$, and $L_y \ge 3$. 
The polynomial $q^2-3q+3$ in (\ref{sqmblamsum}) is $D_4$ in our previous
notation, where 
\beq
D_k = \sum_{s=0}^{k-2}(-1)^s {{k-1}\choose {s}} q^{k-2-s} \ . 
\label{dk}
\eeq

 From the exact solutions for the chromatic polynomials in \cite{wcy}, we have,
 for the cyclic strip of the kagom\'e lattice consisting of a succession of
 hexagons with two interleaved triangles per hexagon of arbitrary length 
(denoted as having width $L_y=2$ in \cite{wcy}) we have 
\beq 
\sum_{j=1}^6 c_{P,kag2,j}\lambda_{P,kag2,j}=q(q-1)^2(q-2)^2 \ .
\label{kag2lamsum}
\eeq
 
  Additional families of graphs are provided by the homeomorphic expansions of
the cyclic and M\"obius $L_y=2$ strips of the square lattice, in which one adds
$k-2$ degree-2 vertices to the upper and lower horizontal edges of each square,
for $k \ge 3$.  These may be viewed as strips of $p$-sided polygons, with
$p=2k$, with each successive pair of polygons sharing one edge.  For $k=2$ and
$k=3$, these families are the $L_y=2$ cyclic and M\"obius strips of the square
and honeycomb lattice, respectively. (In the latter case, the honeycomb lattice
is represented as a brick lattice with the bricks oriented horizontally.)
Following \cite{pg}, we denote the cyclic (cyc) and M\"obius (Mb) strips of 
this type, of length $L_x=m$ as $(Ch)_{k,m,cyc}$ and $(Ch)_{k,m,Mb}$, where 
$Ch$ denotes ``chain''.  Exact solutions for the chromatic polynomials of these
families were given in \cite{pg}. From the construction of 
these graphs, it is clear that $\sum_j c_{P,(Ch)_{k,cyc,j}}=q(q-1)$ and 
$\sum_j c_{P,(Ch)_{k,Mb,j}}=0$ hold, as in eqs. (\ref{ncrelcyc}) and 
(\ref{csummb}).  For the other sums (with $\lambda_{(Ch),k,cyc,j}=
\lambda_{(Ch),k,Mb,j}$), we have
\beq
\sum_{j=1}^4 c_{(Ch)_{k,cyc,j}}\lambda_{(Ch)_{k,cyc,j}}=q^2-3q+1
+2(-1)^{k+1}(q-1)D_{k+1}+D_{2k}
\label{pglamsum}
\eeq
where $D_k$ was defined in (\ref{dk}), and
\beq
\sum_{j=1}^4 c_{(Ch)_{k,Mb,j}}\lambda_{(Ch)_{k,Mb,j}}=-1+2(-1)^k(q-1)D_k+D_{2k}
\ . 
\label{pglamsummb}
\eeq
For $k=2$, the sum (\ref{pglamsum}) is zero, as in (\ref{ttracezero}), and the
sum (\ref{pglamsummb}) has the value $q(q-1)$ as in the $L_y=2$ special case of
(\ref{sqmblamsum}).  As an example of the values for higher--$p$ polygonal
strips, for $k=3$, i.e., the strip of the honeycomb lattice, (\ref{pglamsum})
has the value $q(q-1)^3$ and (\ref{pglamsummb}) has the value $q(q-1)(q-2)^2$.

\vspace{6mm}

For the full Potts model partition functions, from our exact solutions in 
\cite{a,ta} we find 
\beq
\sum_{j=1}^{6} c_{Z,sq2,cyc,j}\lambda_{Z,sq2,cyc,j}=q(v+1)^2(v+q)
\label{sq2cyczlamsum}
\eeq

\beq
\sum_{j=1}^{6} c_{Z,sq2,Mb,j}\lambda_{Z,sq2,Mb,j}=q(v^3+3v^2+3v+q)
\label{sq2mbzlamsum}
\eeq

\beq
\sum_{j=1}^{6} c_{Z,tri2,cyc,j}\lambda_{Z,tri2,cyc,j}=q(v+1)^2(v^2+2v+q)
\label{tri2cyczlamsum}
\eeq

\beq
\sum_{j=1}^{6} c_{Z,tri2,Mb,j}\lambda_{Z,tri2,Mb,j}=q(v+1)(v^3+3v^2+3v+q) \ . 
\label{tri2mbzlamsum}
\eeq

As is evident, for the special case $v=-1$ where the Potts model partition 
function reduces to the chromatic polynomial, these equations reduce to their
analogues for the respective chromatic polynomials. 

In terms of the Tutte polynomials, these formulas read
\beq
\sum_{j=1}^{6} c_{T,sq2,cyc,j}\lambda_{T,sq2,cyc,j}=xy^2
\label{sq2cyczlamsumtut}
\eeq

\beq
\sum_{j=1}^{6} c_{T,sq2,Mb,j}\lambda_{T,sq2,Mb,j}=x+y+y^2
\label{sq2mbzlamsumtut}
\eeq
(which is the Tutte polynomial for the graph known as the ``thick link'' with
three edges, the planar dual to the circuit graph $C_3$)

\beq
\sum_{j=1}^{6} c_{T,tri2,cyc,j}\lambda_{T,tri2,cyc,j}=(x+y)y^2 
\label{tri2cyczlamsumtut}
\eeq
and
\beq
\sum_{j=1}^{6} c_{T,tri2,Mb,j}\lambda_{T,tri2,Mb,j}=y(x+y+y^2) \ . 
\label{tri2mbzlamsumtut}
\eeq

\vfill
\eject
\end{document}